\documentstyle[emulateapj,psfig]{article}

\makeatletter

\newenvironment{inlinefigure}{%
\def\@captype{figure}%
\noindent\begin{minipage}{0.999\linewidth}\begin{center}}
{\end{center}\end{minipage}\smallskip}
\makeatother

\lefthead{Barger et al.}

\begin{document}
\title{Supermassive Black Hole Accretion History Inferred from a Large 
Sample of {\it Chandra} Hard X-ray Sources}
\author{A.\,J.\ Barger,$\!$\altaffilmark{1,2,3,4}
L.\,L.\ Cowie,$\!$\altaffilmark{1,4}
M.\,W.\ Bautz,$\!$\altaffilmark{5}
W.\,N.\ Brandt,$\!$\altaffilmark{6}
G.\,P.\ Garmire,$\!$\altaffilmark{6}
A.\,E.\ Hornschemeier,$\!$\altaffilmark{6}
R.\,J.\ Ivison,$\!$\altaffilmark{7}
F.\,N.\ Owen$\!$\altaffilmark{8}
}

\altaffiltext{1}{Institute for Astronomy, University of Hawaii,
2680 Woodlawn Drive, Honolulu, Hawaii 96822}
\altaffiltext{2}{Department of Astronomy, University of Wisconsin-Madison, 
475 North Charter Street, Madison, WI 53706}
\altaffiltext{3}{Hubble Fellow and Chandra Fellow at Large}
\altaffiltext{4}{Visiting Astronomer, W.\,M.\ Keck Observatory,
jointly operated by the California Institute of Technology and
the University of California}
\altaffiltext{5}{Center for Space Research, Massachusetts Institute
of Technology, Cambridge, MA 02139}
\altaffiltext{6}{Department of Astronomy and Astrophysics,
525 Davey Laboratory, The Pennsylvania State University,
University Park, PA 16802}
\altaffiltext{7}{UK Astronomy Technology Centre, Royal Observatory,
Blackford Hill, Edinburgh EH9 3HJ, U.K.}
\altaffiltext{8}{National Radio Astronomy Observatory, P.O. Box O,
1003 Lopezville Road, Socorro, N.M. 87801}

\slugcomment{Submitted to the Astronomical Journal}

\begin{abstract}
We describe the optical, near-infrared, and radio properties of
a sample of hard ($2-7$~keV) X-ray sources detected in a deep 
{\it Chandra} observation of the field surrounding 
the Abell~370 cluster. We combine these data with similar observations 
of the {\it Chandra} Deep Field-North and the Hawaii 
Survey Field SSA13 to obtain a sample of 69 hard X-ray sources 
(45 are spectroscopically identified) 
with extremely deep 20~cm observations.
We find that about 4\% of the $>L^\ast$ galaxy
population is X-ray luminous at any time and hence that black
hole accretion has a duration of about half a Gyr. 
We find that about 30\% of the summed $2-7$~keV flux
from our total sample is from sources at $z\lesssim1$. 
We estimate the bolometric luminosities of accretion onto
supermassive black holes for our sample, from which we determine maximal
mass inflow rates that increase from $\sim 0.01$~M$_\odot$~yr$^{-1}$ 
at $z\lesssim 0.5$ up to $\sim10$~M$_\odot$~yr$^{-1}$ at $z\gtrsim 1$,
assuming a canonical radiative efficiency $\epsilon\sim 0.1$.
The time history of the {\it accretion rate density} is evaluated;
its maximal integrated value is 
$\rho_{BH}=6\times 10^{-35}$~g~cm$^{-2}$, which is reasonably consistent
with the value inferred from the local black hole mass to bulge mass ratio. 
\end{abstract}


\section{Introduction}
\label{secintro}

X-ray surveys provide a direct probe of active galactic nuclei
(AGN) and hence of
supermassive black hole accretion activity in the Universe.
Soft X-ray samples are biased against sources with high 
line-of-sight absorption, so it is desirable to observe at the
highest possible energies in order to obtain the most complete
sample. Two 1~Ms {\it Chandra} exposures 
(\markcite{brandt01b}Brandt et al.\ 2001b;
P. Rosati, et al., in preparation)
have essentially fully resolved the X-ray background (XRB) 
at about $1.5\times 10^{-15}$~erg~cm$^{-2}$~s$^{-1}$ in the
$2-7$~keV energy band, and the cumulative counts
flatten rapidly below this flux level.
The brighter (100--480~ks) {\it Chandra}
(\markcite{mushotzky00}Mushotzky et al.\ 2000; 
\markcite{giacconi01}Giacconi et al.\ 2001; 
\markcite{horn00}Hornschemeier et al.\ 2000, 2001;
\markcite{brandt01a}Brandt et al.\ 2001a;
\markcite{garmire01}Garmire et al.\ 2001;
\markcite{tozzi01}Tozzi et al.\ 2001) and {\it XMM-Newton}
(e.g., \markcite{hasinger01}Hasinger et al.\ 2001) imaging surveys are
therefore finding nearly all the hard X-ray sources, and 
the analysis of these surveys will provide 
a comprehensive understanding of the nature of the sources
that comprise the XRB at these energies.

The excellent $<1''$ X-ray 
positional accuracy of {\it Chandra} permits the secure
identification of the optical counterparts to the X-ray sources.
One of the most interesting results obtained from optical follow-up 
observations is that almost half of the hard X-ray light arises in 
optically bright galaxies ($I<23.5$) in the $z<1.5$ redshift range.
The other half arises in a mixture of higher redshift AGN and
optically faint galaxies ($I>23.5$), many of whose redshifts likely 
lie in the range $z=1.5$ to 3, based on their colors
(\markcite{crawford01}Crawford et al.\ 2001; 
\markcite{barger01}Barger et al.\ 2001a;
\markcite{cowie01}Cowie et al.\ 2001).

Almost all of the optically bright sources can be spectroscopically 
identified, and most are bulge-dominated galaxies with near-$L^\ast$ 
luminosities.
Contrary to the situation for the faint {\it ROSAT} soft X-ray sources
(\markcite{schmidt98}Schmidt et al.\ 1998), the vast majority ($>85$\%)
do not have broad optical or ultraviolet lines, and about half of
the optically identified sources show no obvious
high ionization signatures of AGN activity.
The {\it Chandra} hard X-ray sources prior to the 1~Ms data are
generally too faint to allow the multiparameter fits to the
X-ray spectra that would directly measure the intrinsic column densities.
However, the soft to hard X-ray flux ratios suggest that
the sources are highly absorbed systems whose
high column densities could effectively extinguish the optical,
ultraviolet, and near-infrared continua from the AGN and render
traditional identification techniques ineffective.

One of the most exciting avenues in the study of the hard X-ray
source populations is determining the times and duration of 
distant supermassive black hole activity. 
\markcite{barger01}Barger et al.\ (2001a) found 
that about 7\% of optically luminous galaxies are X-ray active at 
any time. A similar active fraction (6\%) was later found by 
\markcite{horn01}Hornschemeier et al.\ (2001) for the 225~ks
{\it Chandra} Deep Field-North (CDF-N) data.
Thus, these studies suggest that, on average, accretion activity lasts 
for on order of half a Gyr over the lifetime of a galaxy.

In this paper we present a multiwavelength study of the hard X-ray 
sources detected in the field containing the $z=0.37$ massive 
lensing cluster A370 (\markcite{bautz00}Bautz et al.\ 2000),
the SSA13 field (\markcite{mushotzky00}Mushotzky et al.\ 2000;
\markcite{barger01}Barger et al.\ 2001a, hereafter B01), and
the CDF-N field (\markcite{horn01}Hornschemeier et al.\ 2001, 
hereafter H01).

\markcite{bautz00}Bautz et al.\ (2000) presented the
detection in hard X-rays of the two distant submillimeter sources
at $z=2.80$ and $z=1.06$ behind the A370 cluster; these sources are
gravitationally amplified by factors of 2.4 and 2.1, respectively. 
The remaining 13 hard X-ray sources are not strongly 
affected either by the cluster lens (amplifications between 1 and 1.3) 
or by the diffuse emission from the cluster.
In the following analysis we take into account the two large
amplifications but neglect the smaller amplifications, treating
the remaining objects as a field sample.

When we combine the A370 data with the SSA13 and CDF-N data, we
obtain a sample of 69 hard X-ray selected sources with extremely deep 
20~cm and optical imaging data. We have spectroscopic redshifts
for 45 of the 69 sources. 
We use our combined sample to determine the optical properties 
of the X-ray sources that comprise the hard XRB, the duty cycle of 
X-ray activity, and the luminosities associated with AGN accretion.  
We then evaluate the mass inflow rate onto the supermassive black holes 
and deduce the time history of the accretion rate density.

We take $H_o=65\ h_{65}$~km~s$^{-1}$~Mpc$^{-1}$ and use an
$\Omega_{\rm M}=1/3$, $\Omega_\Lambda=2/3$ cosmology
throughout.

\section{A370 Sample and Observations}
\label{secdata}

\subsection{X-ray Observations}
\label{secxray}

{\it Chandra} observed A370 on 22--23 October 1999 for a total of 
93.7~ks with the ACIS S3 detector.  The telescope optical axis was 
pointed approximately $2.4'$~E of the cluster center. The ACIS 
focal plane temperature during the observation was $-110^\circ$~C.
The detectors were read out in timed-exposure mode and events were
processed on board in faint mode. The (re)processed data were delivered 
by the {\it Chandra} X-ray Observatory Center (CXC) on 4~September 2000. 

We processed the Level 2 CXC data products using standard methods.
Periods of background flaring were removed using the tool {\tt lc\_clean}
devloped by Maxim Markevitch. 
The total S3 counting rate (0.1 to 10~keV) was measured in 100~s 
time bins, and a quiescent background of 1.59~counts~s$^{-1}$
(over the full S3 detector) was estimated from one 17~ks period.
Time bins in which the observed background differed from the quiescent 
value by more than 2.3~$\sigma$ were excluded from our subsequent analysis. 
The net flare-free exposure time was 63.2~ks. 

We restrict attention to events occurring in the S3 detector with 
{\it ASCA} grades 0, 2, 3, 4, and 6 and with energies between 0.3 
and 7~keV in order to maximize the signal-to-noise ratio.
The $0.3-7$~keV background level in source-free regions of the field is 
$(2.09 \pm 0.02) \times 10^{-6}$~counts~s$^{-1}$~arcsec$^{-2}$. 

We used the CIAO tool {\tt wavdetect}, with a significance parameter value
of $10^{-7}$, to search for sources in the $0.5-7$~keV, $0.5-2$~keV, and 
$2-7$~keV bands. A total of 34, 25, and 17 sources, respectively, 
were detected in each band. The expected false detection rate in each 
band is 0.1~source. We selected all sources that were more than
$15''$ from the chip edges; this removed one soft and hard band 
source from the {\tt wavdetect} sample. The individual soft and 
hard band sources 
were then reviewed and compared with the results of a cell detection 
method. Two hard band sources detected by {\tt wavdetect} had 
only 5 counts each. One of these sources could not be reproduced 
with the cell detection method and was removed. 
The other source was measured to have 8.3 counts in a
$2''$ radius aperture. For this source we have used the counts
determined by the aperture photometry; it is marked in Table~1.
Subsequent work on the 1~Ms CDF-N data has shown that {\tt wavdetect} 
detections with a significance parameter of $10^{-7}$
are highly reliable (see \S3.2.3 of \markcite{brandt01b}Brandt et al. 2001b), 
but, for the present paper, we have chosen to remain with our
initial sample selection.
We are left with a total of 15 hard X-ray sources in our A370
sample.

We converted the detected counts for each source into fluxes 
assuming a photon spectral flux distribution function 
$\frac{dN}{dE} \propto E^{-\Gamma}$, 
modified by the expected Galactic absorbing column density of 
$4.0\times 10^{20}$~cm$^{-2}$, assuming an intrinsic $\Gamma=2$.
In creating exposure maps to account for energy-dependent vignetting 
and detector efficiency variations using the CIAO tools {\tt mkinstmap} 
and {\tt mkexpmap}, we also assumed $\Gamma=2$. Since the fluxes
in the SSA13 and CDF-N fields were analyzed with a
harder $\Gamma$ ($\Gamma=1.2$ was the counts-weighted mean photon
index for the SSA13 sample, and $\Gamma=1.4$ was assumed for sources 
in the CDF-N sample when the individual $\Gamma$s could not be
determined), we repeated the counts-to-flux conversion procedure 
on our A370 sample using $\Gamma=1.2$. We found that soft band fluxes 
increased by 25\% and hard band fluxes increased by 33\%.
To be more consistent with the SSA13 and CDF-N samples, we
adopt these correction factors in our subsequent analysis.

Ten of the 34 sources detected in the broad-band image
with {\tt wavdetect} coincide within $2''$ with radio 
sources detected at high significance with the VLA (see \S\ref{secradio}).
The mean position offsets ({\it Chandra} -- VLA) for these sources 
are $0.4''$ and $0.0''$ in right ascension and declination, 
respectively. The dispersion of the positions is $0.55''$.
We have corrected the positions in the tables to allow for
this small offset.

\begin{figure*}[tb]
\centerline{\psfig{figure=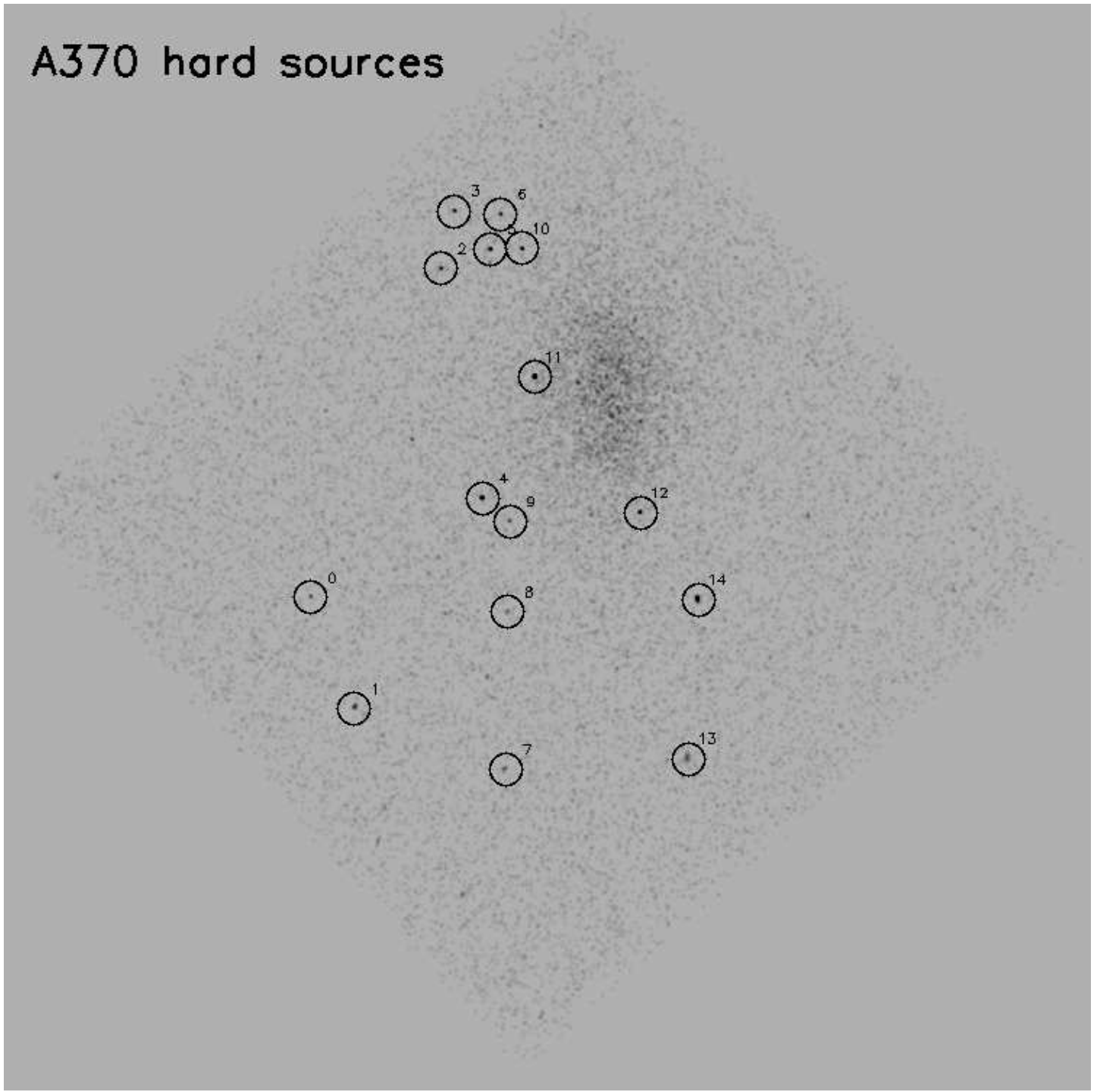,angle=90}}
\figurenum{1}
\figcaption[]{
{\it Chandra} hard X-ray image of the A370 field.
The 15 significant hard X-ray source positions are identified by
the small circles. The X-ray image has been smoothed with
a $2.5''$ boxcar smoothing to allow the X-ray sources to be clearly seen.
\label{fig1}
}
\end{figure*}

Figure~\ref{fig1} shows the {\it Chandra} $2-7$~keV image of A370 
with the final 15 hard X-ray source positions identified by small 
circles.  Table~\ref{tab1} details these 15 sources,
ordered by increasing RA. The first five columns 
in Table~\ref{tab1} include the source identification,
RA(2000), Dec(2000), $2-7$~keV flux, and $0.5-2$~keV flux.
Table~\ref{tab2} details the supplementary 14 sources in our
field that were not detected in the hard band but were 
detected in the soft band. The first four columns include the 
source identification, RA(2000), Dec(2000), and $0.5-2$~keV flux.
The remaining entries in the tables are discussed in subsequent
sections.

\subsection{Optical and Near-Infrared Imaging}
\label{secopt}

Deep multicolor images ($V$, $R$, $I$) that cover the A370 
{\it Chandra} field were obtained using the Low-Resolution Imaging 
Spectrometer (LRIS; Oke et al.\ 1995) on the Keck 10~m telescopes.
Wide-field near-infrared images were obtained with the UH~2.2~m
telescope using the University of Hawaii Quick Infrared Camera 
(QUIRC; \markcite{hodapp96}Hodapp et al.\ 1996) and a 
notched $HK'$ ($1.9\pm 0.4$~$\mu$m) filter. Details of the observations 
can be found in \markcite{cowie01}Cowie et al. (2001) and 
E. Hu, L. Cowie, \& R. McMahon, in preparation.

\begin{figure*}[tb]
\centerline{\psfig{figure=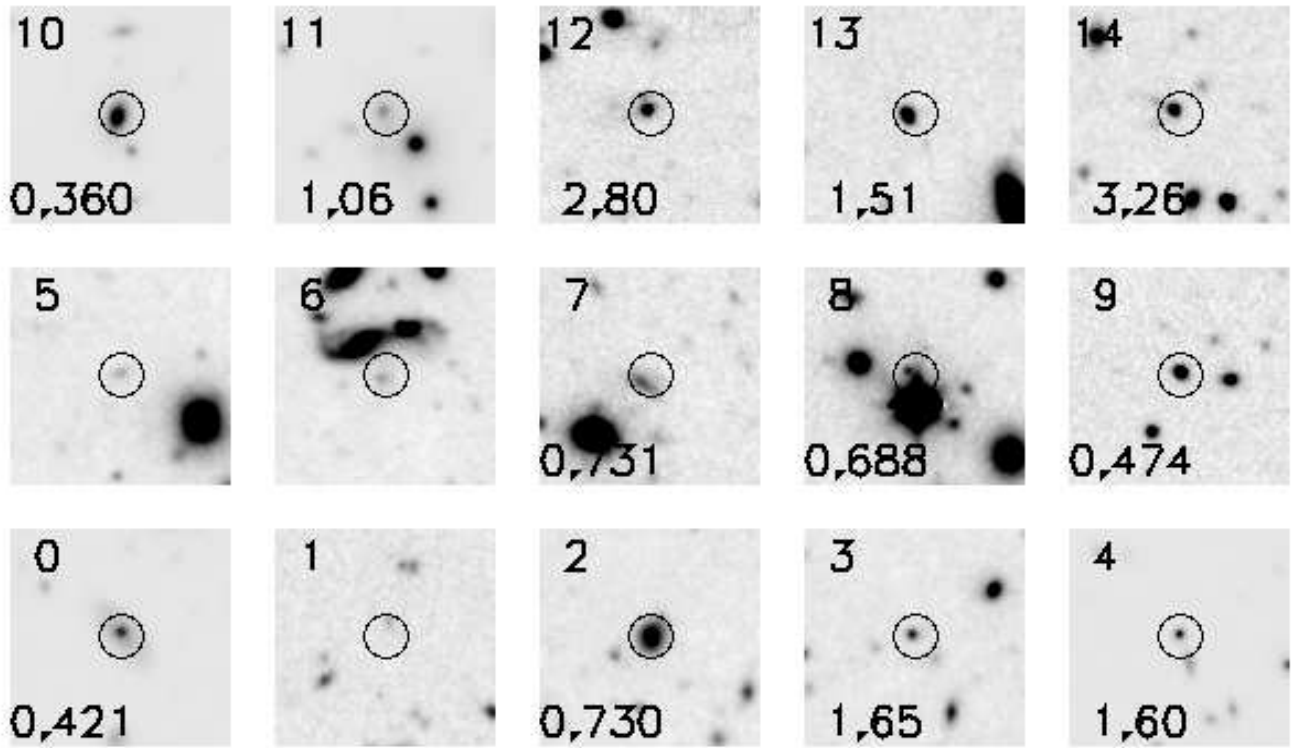,angle=90,width=7.5in}}
\figurenum{2}
\figcaption[]{
$9''\times 9''$ $I$-band thumbnail
images of the A370 hard X-ray sources listed in Table~1.
The images are from ultradeep data obtained with LRIS on Keck.
The identification numbers are as in Table~1.
A circle of $1.5''$ radius, typical
of the maximum positional uncertainty, is superimposed on
each thumbnail. Redshifts, where available, are given in the lower
left-hand corners of the thumbnails. North is up and East is to the left.
For the more luminous objects we have shown the images at higher
surface brightness to allow the positions of the objects with
respect to the centers of the galaxies to be seen.
\label{fig2}
}
\end{figure*}

Figure~\ref{fig2} shows thumbnail $I$-band images of the 15 hard
X-ray sources listed in Table~\ref{tab1}. In selecting the 
optical counterparts, we considered only
sources within a $1.5''$ radius of the nominal X-ray position. 
The optical separations are given in column~11 of Table~1.
The magnitudes were measured in $3''$ diameter apertures
at the optical center and corrected to approximate total
magnitudes using an average offset 
(\markcite{cowie94}Cowie et al.\ 1994); henceforth, we refer to
these as corrected $3''$ diameter apertures.
The $HK'$, $I$, $R$, and $V$ magnitudes are given in columns
6--9 of Table~1. The $2\sigma$ limits are approximately 
$HK'=20.9$, $I=24.9$, $R=26.1$, and $V=26.1$. All of the
sources have identifications in the $I$ band, and none of the 
counterparts are fainter than $I=23.5$, which was found to be the 
effective optical limit for spectroscopic identification of the 
sources in the SSA13 hard X-ray sample (B01).

\subsection{Keck Spectroscopy}
\label{seckeck}

We made spectroscopic observations of the X-ray sources in 
the A370 field using LRIS slit-masks
on the Keck 10~m telescope on UT 2000 February 6 and 
2001 January 23--24. We positioned the slit at the center of 
the optical counterpart. We used $1.4''$ wide slits. 
In the February run we used the 300 lines~mm$^{-1}$ grating 
blazed at 5000~\AA, which gives a wavelength resolution of 
approximately 16~\AA\ and a wavelength coverage of 
approximately 5000~\AA. The wavelength range for each object 
depends on the exact location of the slit in the mask but 
is generally between $\sim5000$ and 10000~\AA. 
In the January run we used either the 400~lines~mm$^{-1}$ grating 
blazed at 8500~\AA, which gives a wavelength resolution of
approximately 12~\AA\ and a wavelength coverage of
approximately 4000~\AA, or the 600~lines~mm$^{-1}$ grating blazed
at 10000~\AA, which gives a wavelength resolution of
approximately 8~\AA\ and a wavelength coverage of
approximately 2500~\AA. We used the D560 dichroic to obtain 
simultaneous LRIS-B observations with the 600~lines~mm$^{-1}$ 
grism blazed at 5000~\AA, which for our slit width
gives a wavelength resolution of approximately 7~\AA\ and a 
wavelength coverage of approximately 2000~\AA.

Each slit mask was observed for 1.5\ hr, broken into three sets of 
0.5\ hr exposures. Fainter objects were put in more than one mask
to provide longer total exposures.
Conditions were photometric with seeing $\sim 0.6''-0.8''$~FWHM.
The objects were stepped along the slit by $2''$ in each
direction, and the sky backgrounds were removed using the median of
the images to avoid the difficult and time-consuming problems of
flat-fielding LRIS data. Details of the spectroscopic reduction
procedures can be found in \markcite{cowie96}Cowie et al.\ (1996).

\begin{figure*}[tb]
\centerline{\psfig{figure=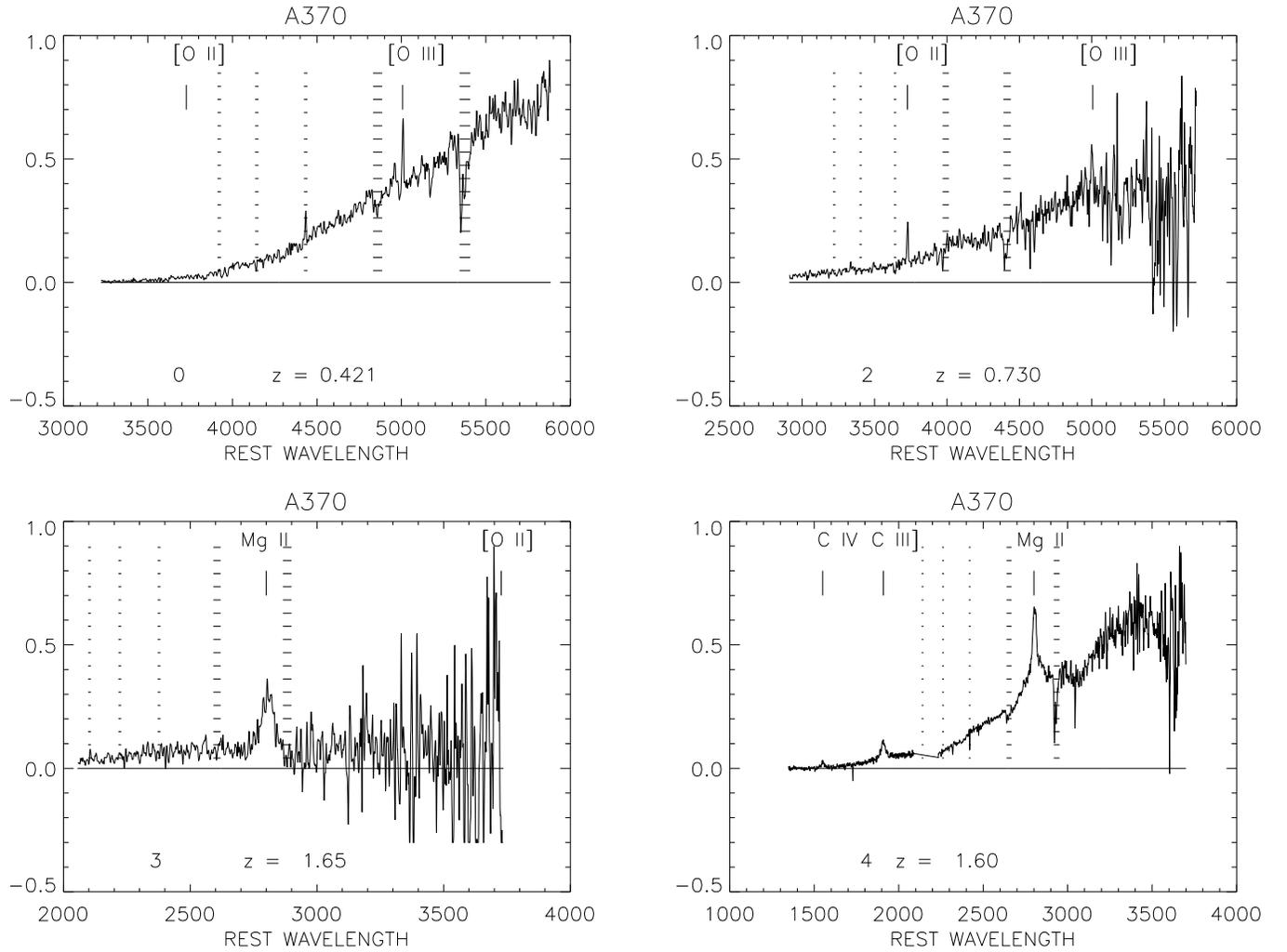,angle=90,width=7.5in}}
\figurenum{3}
\figcaption[]{
Keck LRIS spectra of the A370 hard X-ray sources with secure redshift
identifications (except source 12). 
The spectra are shown in the rest frame with
source number and the measured redshift at the bottom of the frame.
Dashed lines mark strong atmospheric absorption or emission features.
\label{fig3}
}
\end{figure*}

We successfully obtained redshift identifications for 10 of the
15 hard X-ray sources in our sample. Two others were already 
identified in the literature:  source 12 from 
\markcite{ivison98}Ivison et al.\ (1998)
and source 11 from \markcite{barger99b}Barger et al.\ (1999b) 
and \markcite{soucail97}Soucail et al.\ (1999). The spectra
for the hard X-ray sources (except source 12) 
are shown in Fig.~\ref{fig3}, and the 
redshifts are given in column~10 of Table~\ref{tab1}. 
Strong emission line features are marked on the plots.
We also obtained redshift identifications for 8 of the 14 additional  
soft X-ray sources in the field, and these are given in column~5 
of Table~\ref{tab2}.

\subsection{Radio Observations}
\label{secradio}

A very deep 1.4~GHz map of the A370 field was
obtained by F. Owen, et al., in preparation, with the
National Radio Astronomy Observatory's Very Large
Array (VLA). The data comprise 45~hr in A configuration and 20~hr in B
configuration, using the spectral-line correlator mode together with short
integration times to eliminate bandwidth smearing.
A large field-of-view ($40'$ in diameter) was completely imaged.
In the region covered by the {\it Chandra} results reported here, the effective
resolution is $1.6$ arcseconds and the rms noise is typically $5.5-6.5\mu$Jy,
depending on details of the residual sidelobe distribution from nearby
sources; thus, we have adopted 20~$\mu$Jy as a 3$\sigma$ upper limit for 
undetected sources. The radio fluxes were measured at the highest
peak within $2''$ of each X-ray position; these fluxes are given in column~12 
of Table~\ref{tab1} and column~6 of Table~\ref{tab2}. 

%
%
\begin{inlinefigure}
\psfig{figure=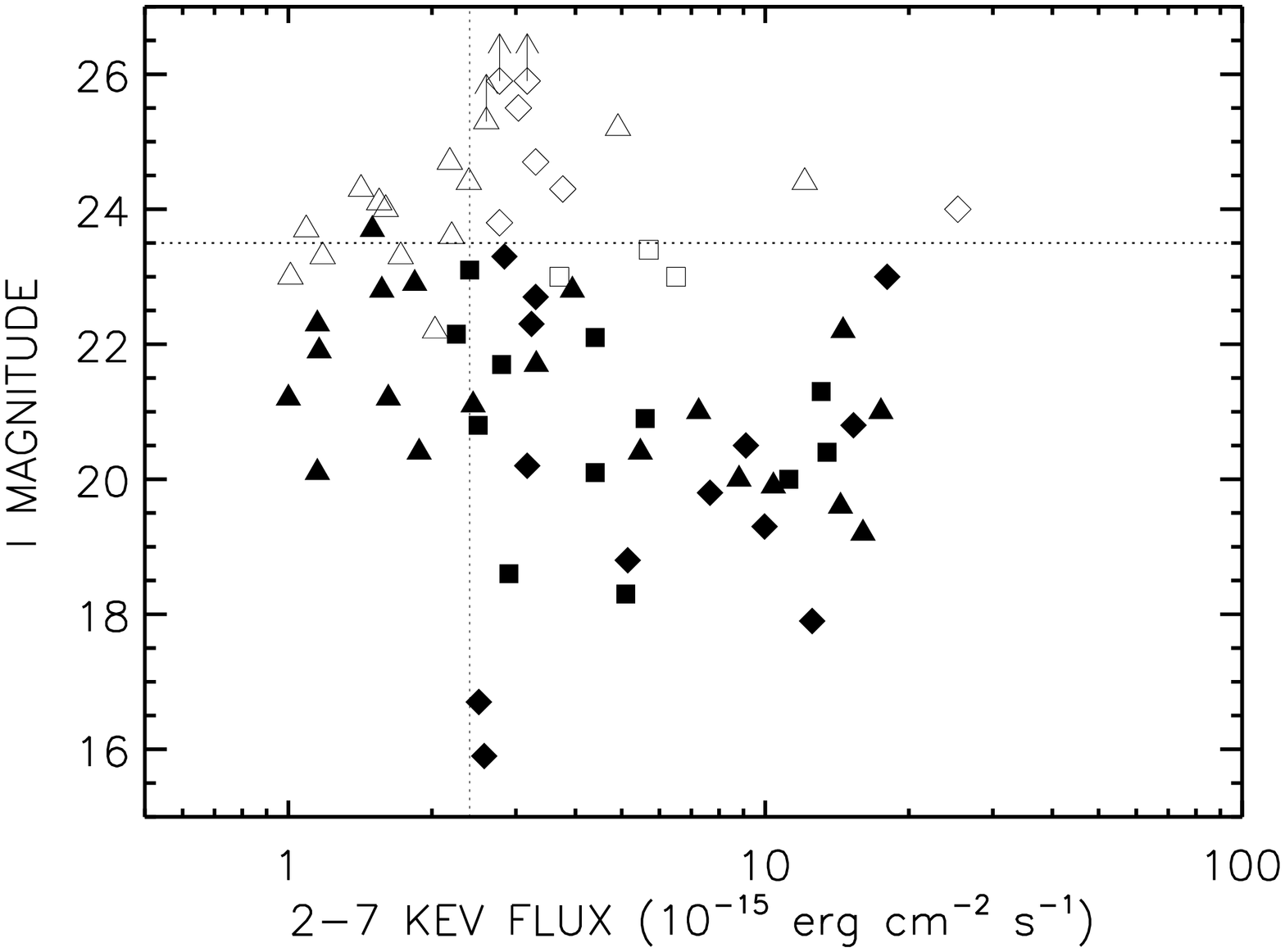,angle=90,width=3.5in}
\vspace{6pt}
\figurenum{4}
\caption{
$I$ magnitude versus $2-7$~keV flux for the hard X-ray
sources in the A370 (squares), CDF-N (triangles), and SSA13 (diamonds)
fields. Sources with spectroscopic
identifications are denoted by filled symbols.
The horizontal dotted line at $I=23.5$ indicates
our magnitude division between optically faint and
optically bright galaxies; all but seven of the sources
brighter than this limit have spectroscopic
identifications. The dotted vertical line at
$2.4\times 10^{-15}$~erg~cm$^{-2}$~s$^{-1}$ is
the hard X-ray detection limit for the A370 and the SSA13
samples. The hard X-ray flux of one of the sources in the
A370 field (source 12) is fainter than the hard X-ray
detection limit after correcting for cluster magnification.
}
\label{fig4}
\addtolength{\baselineskip}{10pt}
\end{inlinefigure}

\section{Combined A370, CDF-N, and SSA13 Datasets}

We hereafter use a combined hard X-ray sample from
the A370, CDF-N, and SSA13 fields. In order to merge the samples
we converted the $2-10$~keV fluxes from B01
and the $2-8$~keV fluxes from H01 to the $2-7$~keV band using
$\Gamma=1.2$. The X-ray detection limit for the A370 and SSA13
samples is $2.4\times 10^{-15}$~erg~cm$^{-2}$~s$^{-1}$ ($2-7$~keV),
and the detection limit for the CDF-N sample is
$5.6\times 10^{-16}$~erg~cm$^{-2}$~s$^{-1}$ ($2-7$~keV).
For the present work we have restricted the CDF-N sample
to sources with fluxes greater than
$1.0\times 10^{-15}$~erg~cm$^{-2}$~s$^{-1}$ ($2-7$~keV) to
provide a conservative sample with uniform flux selection over
the field area.
This CDF-N sample, adapted from H01, is given in Table~\ref{tab3}. We
include in the table $HK'$ and $B$-band magnitudes from
\markcite{barger99a}Barger et al.\ (1999a)
and A. Barger, et al., in preparation, respectively.
We also include 20~cm data from
\markcite{richards00}Richards et al.\ (2000).
The combined sample consists of 69 sources with $2-7$~keV fluxes
ranging from $1.0\times 10^{-15}$~erg~cm$^{-2}$~s$^{-1}$ to
$2.5\times 10^{-14}$~erg~cm$^{-2}$~s$^{-1}$.

%
%

\begin{inlinefigure}
\psfig{figure=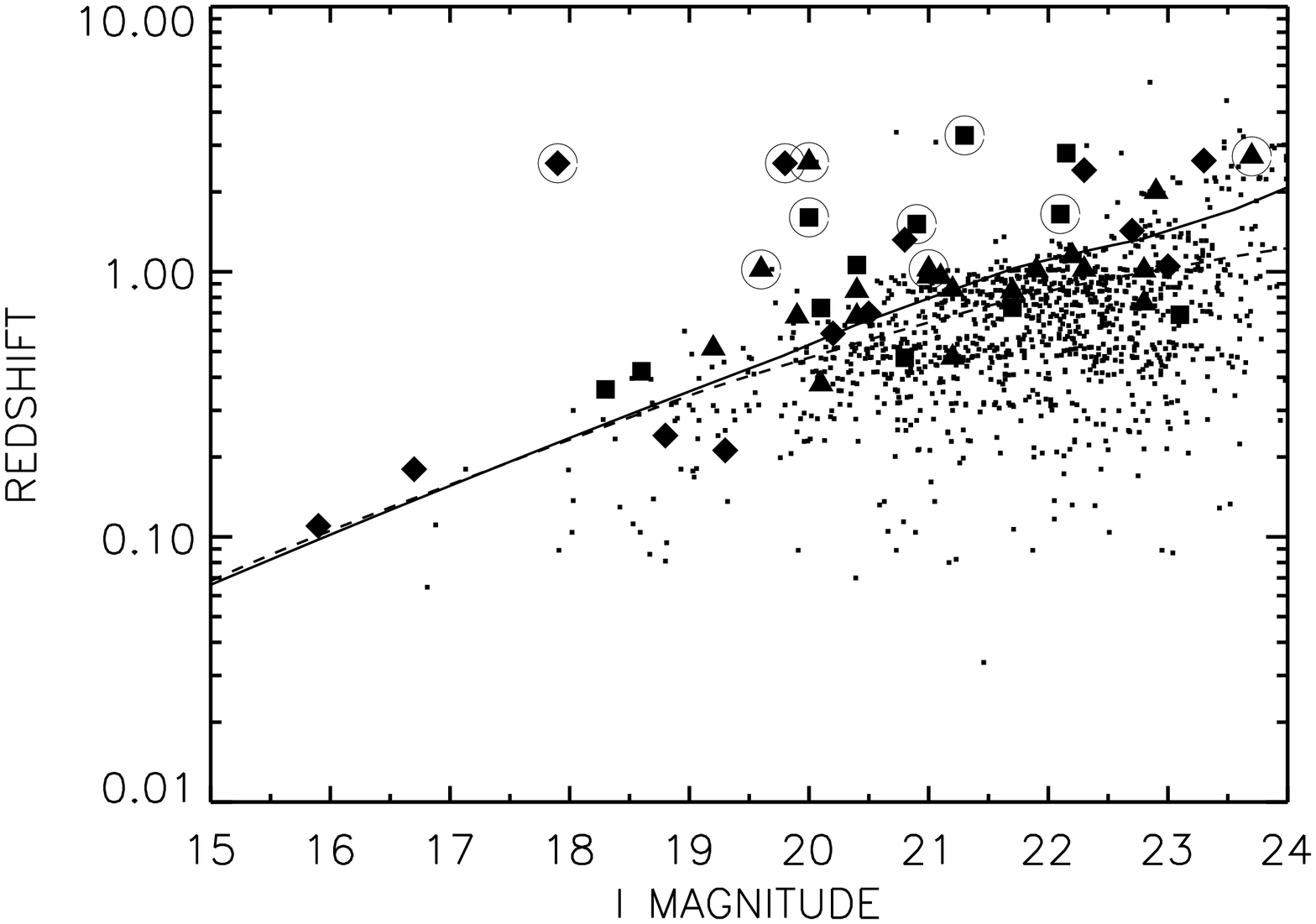,angle=90,width=3.5in}
\vspace{6pt}
\figurenum{5}
\caption{
Redshift versus $I$ magnitude for the 45 hard X-ray sources
with spectroscopic identifications in the A370 (squares),
CDF-N (triangles), and SSA13 (diamonds) fields and for an $I<24$ field
galaxy sample from the CDF-N, SSA13, and SSA22 (small symbols) fields.
Superimposed are Coleman, Wu, \& Weedman (1980) evolved tracks for an
early-type galaxy (solid) and an irregular galaxy (dashed) with
$M_I=-22.5$. Sources with broad emission lines are circled.
}
\label{fig5}
\addtolength{\baselineskip}{10pt}
\end{inlinefigure}

\section{Optical Properties of the Combined Hard X-ray Samples}

\subsection{Magnitudes}
\label{secmag}

In Fig.~\ref{fig4} we plot $I$ magnitude versus $2-7$~keV
flux for the combined sample. The filled symbols
denote sources with spectroscopic identifications;
45 of the 69 sources (65\%) have redshift identifications, and
all but seven of the sources brighter than $I=23.5$ (as indicated
by the horizontal line) have redshifts. The three samples are
very similar in their distribution of optical magnitudes
and redshift identifications.

The median and mean $I$ magnitudes as a function of hard
X-ray flux are summarized in Table~4. As has been noted
previously (e.g., H01), the data are consistent with a constant
optical to hard X-ray ratio, though there is a very wide range of
optical magnitudes for a given X-ray flux.

In Fig.~\ref{fig5} we plot redshift versus $I$ magnitude
for the 45 spectroscopically identified hard X-ray sources in
the combined sample (large symbols). For comparison, we
include on the figure an optically selected $I<24$ field galaxy 
sample (small symbols). We superimpose on the figure tracks calculated 
from the spectral energy distributions (SEDs)
of \markcite{coleman80}Coleman, Wu, \& Weedman (1980) for an 
early-type galaxy (solid) and
an irregular galaxy (dashed) in the absence of evolution, both with absolute magnitude
$M_I=-22.5$ in the assumed cosmology.
B01 noted that the hard X-ray sources predominantly lie in the
most optically luminous galaxies, and as Fig.~\ref{fig5} shows,
this is also the case for the combined sample.
\markcite{brandt01a}Brandt et al.\ (2001a) have recently noted that this
property appears to continue to hold to fainter X-ray flux levels,
$\sim 3\times 10^{-16}$~erg~cm$^{-2}$~s$^{-1}$.
We have circled the sources with broad emission lines in their spectra. 
These sources tend to lie to the left of the evolved galaxy tracks for a 
given redshift, indicating substantial optical brightening due to the 
AGN's optical emission.

%
%

\begin{inlinefigure}
\psfig{figure=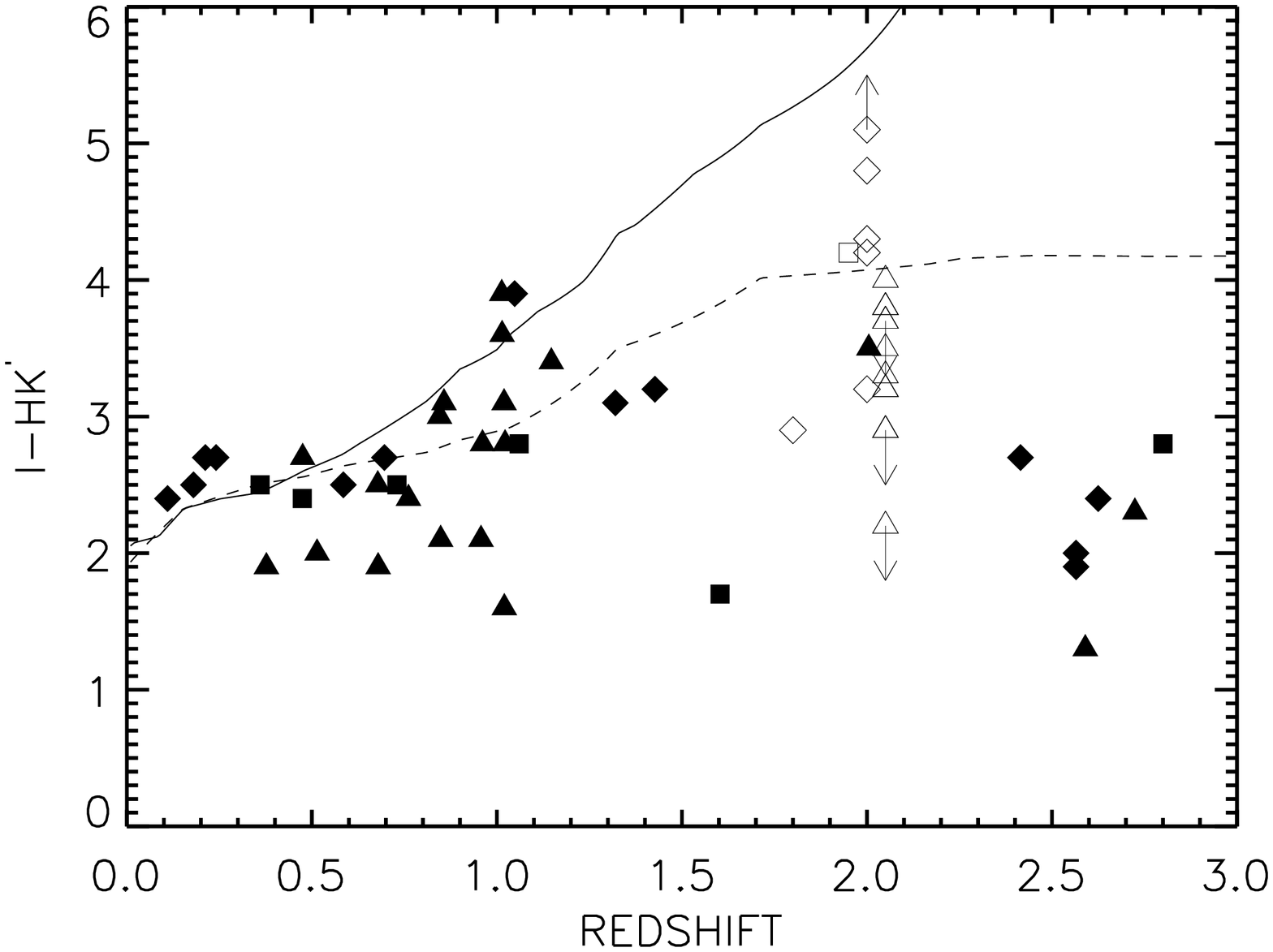,angle=90,width=3.5in}
\psfig{figure=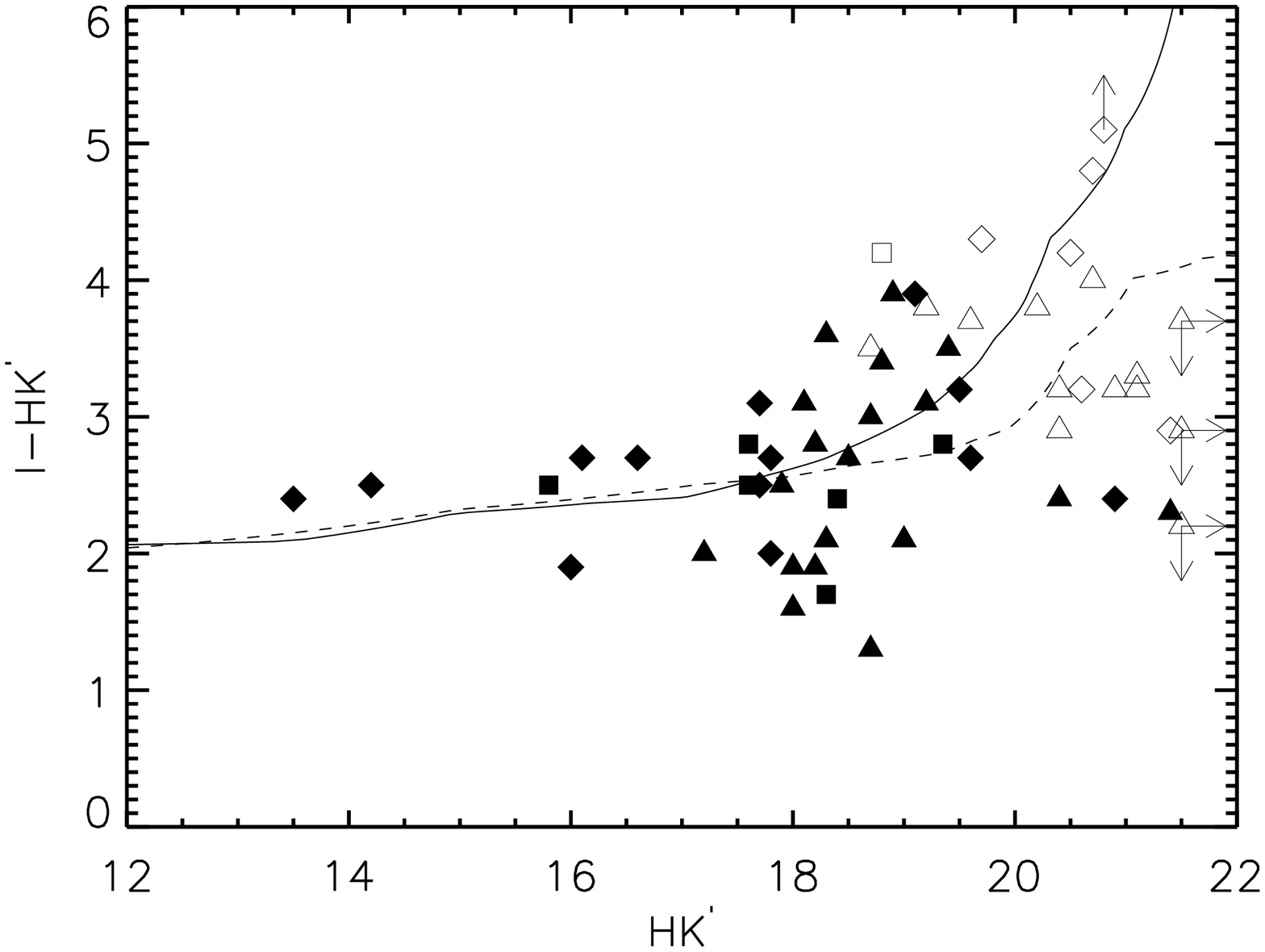,angle=90,width=3.5in}
\vspace{6pt}
\figurenum{6}
\caption{
(a)~$I-HK'$ color versus redshift and (b)~$I-HK'$ color versus
$HK'$ magnitude for the hard X-ray sources in the A370 (squares),
CDF-N (triangles), and SSA13 (diamonds) fields with optical
and/or near-infrared detections (two sources --- one in the
CDF-N field and one in the SSA13 field --- with only limits on
both magnitudes are not shown). The spectroscopically unidentified
sources are nominally placed at $z=2$ (slight redshift offsets have
been applied to allow the three fields to be distinguished).
The source from the SSA13 field that has a millimetric redshift of 1.8
(as deduced from the submillimeter to radio flux ratio; see B01)
is shown as an open diamond in (a) at $z=1.8$.
The overlays are Coleman, Wu, \& Weedman (1980)
tracks for an early-type galaxy (solid curve) and a
spiral galaxy (dashed curve) with $M_{HK'}=-25.0$.
}
\label{fig6}
\addtolength{\baselineskip}{10pt}
\end{inlinefigure}

\subsection{Optical Spectral Classification}

Following B01, we classify the optical spectra of the hard X-ray sources
into three general categories: (i)~broad emission lines,
(ii)~clear signs of [NeIII] and/or [NeV] and/or CIV, and
(iii)~no sign of any of the above signatures.
The spectra are generally of high quality, and strong high ionization
lines are easily seen.
We have included four new redshifts in the CDF-N sample
(sources 9, 13, 15, and 30); these spectra will be presented in
a forthcoming paper (A. Barger, et al., in preparation).
We measured all but three of the redshifts in Table~\ref{tab3}
from our own Keck LRIS spectra. The redshifts of the three sources
in our table for which we do not have Keck spectra
(sources 22, 25, and 26) are taken from the compilation in H01
(the first two were measured by H01 from Hobby-Eberly Telescope
spectra, and the last one was measured by
\markcite{cohen00}Cohen et al.\ (2000) from a Keck LRIS spectrum.)
Table~\ref{tab5} gives the numbers of sources in each of the three
optical spectral classes for each sample.
The identified fractions across the samples are divided fairly evenly
into sources that show broad or high ionization lines in their optical
spectra and sources that show no such optical signatures.

\subsection{Colors}

In Fig.~\ref{fig6}a we plot $I-HK'$ color versus redshift
for sources in the combined sample with measured $HK'$ and $I$
magnitudes or limits and spectroscopic identifications (filled symbols).
We place sources without spectroscopic identifications at
$z=2$ and identify them with open symbols.
In Fig.~\ref{fig6}b we plot $I-HK'$ versus $HK'$ for sources
with measured $HK'$ and $I$ magnitudes or limits.
The overlays are \markcite{coleman80}Coleman, Wu, \& Weedman (1980)
tracks for a non-evolving early-type galaxy (solid curve) and
a spiral galaxy (dashed curve) with $M_{HK'}=-25.0$.
The colors of most of the spectroscopically identified $z<1.5$
galaxies are in the range of the galaxy tracks. A sizeable
fraction of the spectroscopically unidentified sources have colors
that are consistent with luminous galaxies at $z>1.5$.

%
%

\begin{inlinefigure}
\psfig{figure=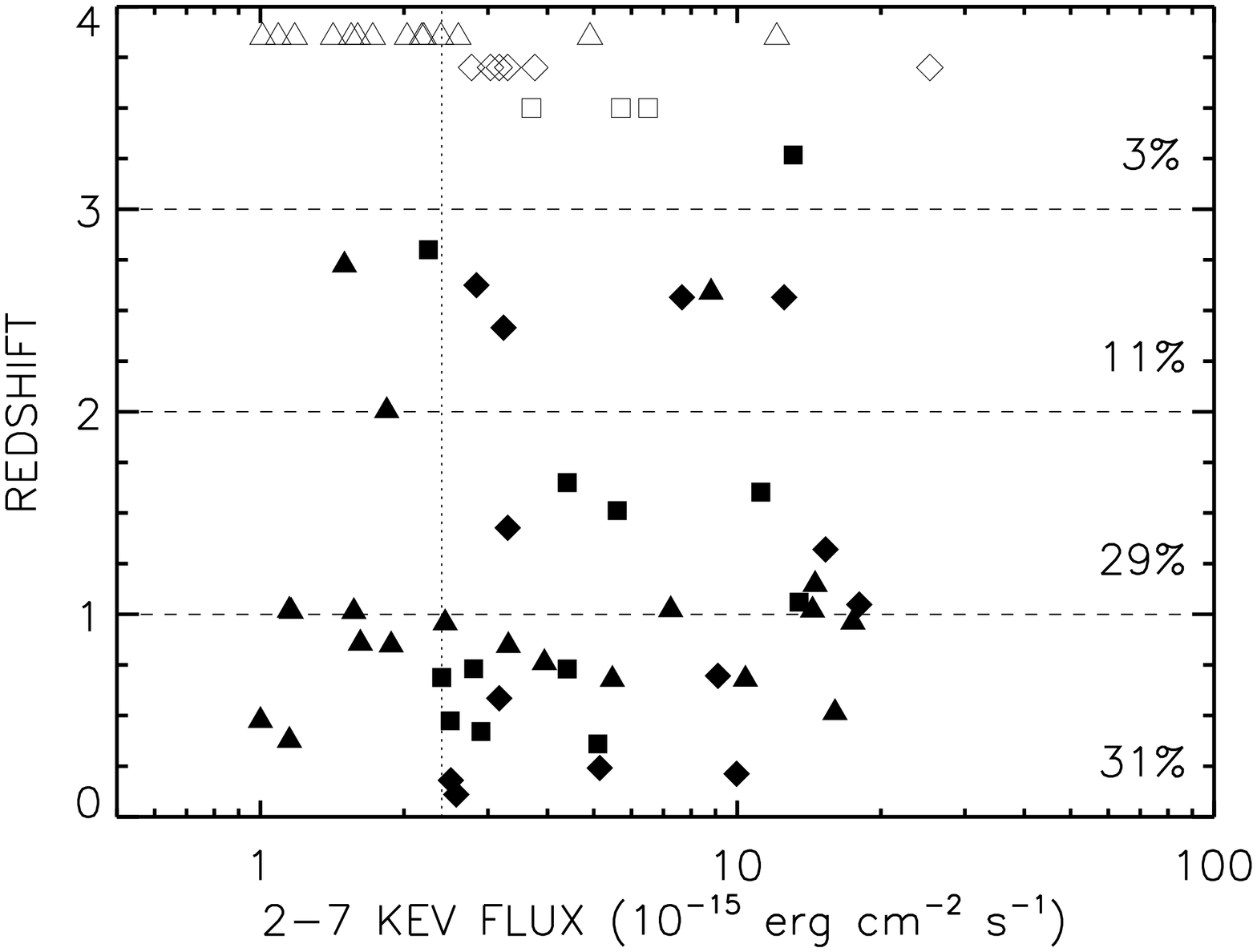,angle=90,width=3.5in}
\vspace{6pt}
\figurenum{7}
\caption{
Redshift versus $2-7$~keV flux for the hard X-ray sources
in the A370 (squares), CDF-N (triangles), and SSA13 (diamonds)
fields. Sources with spectroscopic identifications are denoted by
filled symbols. The dotted vertical line at
$2.4\times 10^{-15}$\ erg\ cm$^{-2}$\ s$^{-1}$ is
the hard X-ray detection limit for the A370 and SSA13 samples.
The dashed horizontal lines separate the sources into redshift bins.
The percentages indicate how much hard X-ray flux originates in
each redshift bin relative to the total flux in our sample.
Spectroscopically unidentified sources are arbitrarily plotted at 
the top of the figure and are not included in the listed percentage 
contributions; many are likely to lie in the redshift range $z=1.5$ to 3.
}
\label{fig7}
\addtolength{\baselineskip}{10pt}
\end{inlinefigure}

\subsection{Redshift Distribution}

We show the redshift distribution of the sources versus
hard X-ray flux in Fig.~\ref{fig7}, where we also give the
percentage of the hard X-ray flux originating in each redshift
interval relative to the total flux in our sample.
Figure~\ref{fig7} shows that about 30\% of
the summed hard X-ray flux from all of our sources has occurred
since $z=1$, when the Universe was half its present age.
The spectroscopically unidentified sources (which are most likely
to lie in the redshift range $z=1.5$ to $z=3$)
are not included in the percentages given,
but the spectroscopically identified
sources already make up about three-quarters of the summed hard X-ray
flux in our sample.

\section{The Duty Cycle of X-ray Activity}
\label{dutycycle}

The {\it Chandra} data allow us to
estimate the duty cycle of X-ray activity in galaxies.
Spectroscopically identified hard X-ray sources are common in 
bulge-dominated (see thumbnail images in Fig.~\ref{fig2};
see also Fig.~2 of B01 and Fig.~6 of H01)
optically luminous galaxies (see Fig.~\ref{fig5}).
In B01 we measured the X-ray properties of known luminous
galaxies in an optically selected field galaxy sample and found
that $7^{+5}_{-3}$\% were X-ray active.

%
%

\begin{inlinefigure}
\psfig{figure=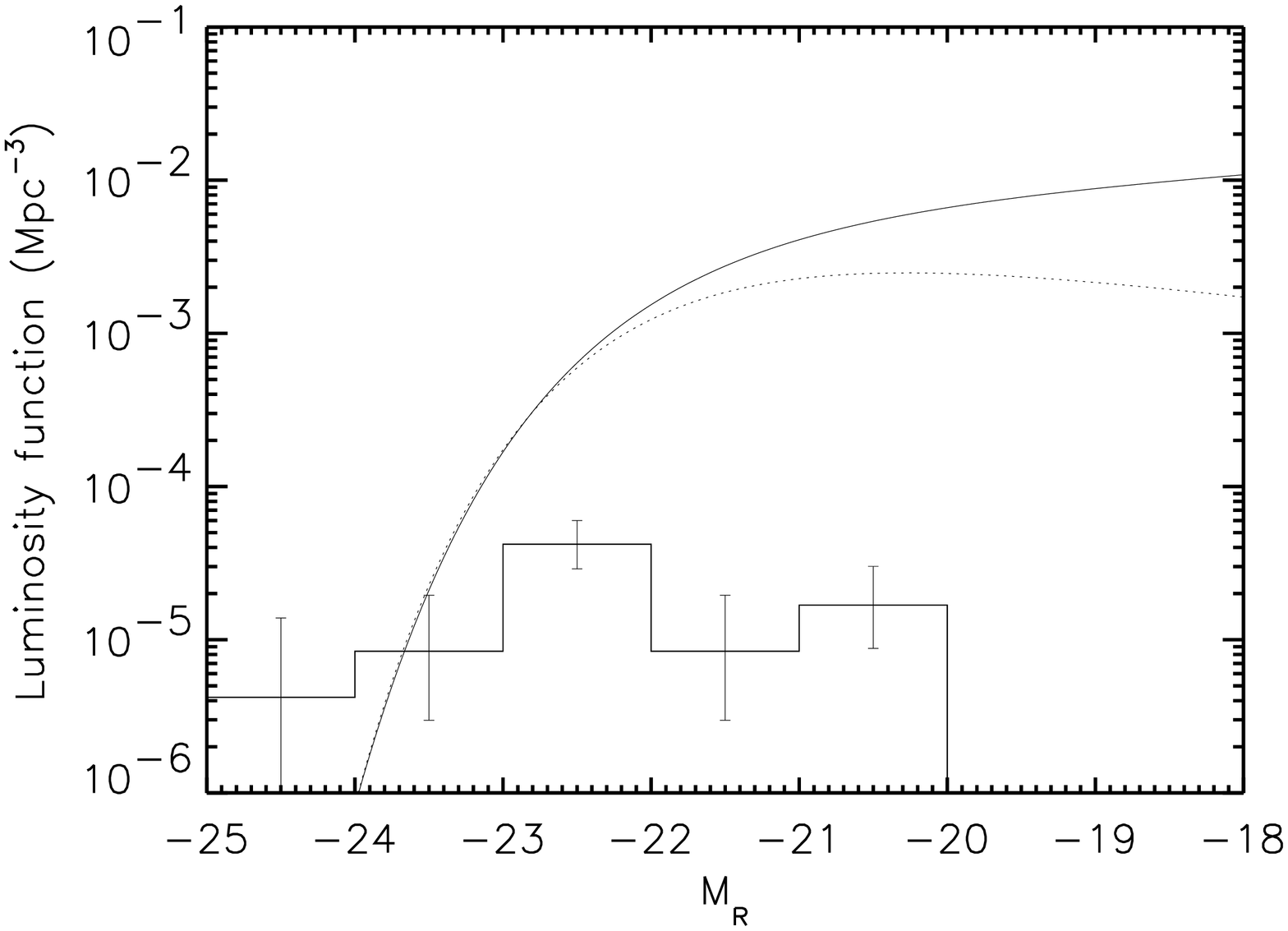,angle=90,width=3.5in}
\vspace{6pt}
\figurenum{8}
\caption{
$R$-band luminosity function from the $z=0-1$ combined hard X-ray
selected sample (histogram; uncertainties are $1\sigma$).
The local $R$-band luminosity functions of
Lin et al.\ (1996; dotted line) and Geller et al.\ (1997; solid line)
are shown for comparison.
}
\label{fig8}
\addtolength{\baselineskip}{10pt}
\end{inlinefigure}

Here we adopt a different approach: we construct the
optical luminosity function (LF) of our X-ray selected sample
and compare it with the local LF.
The optical LF of red galaxies is relatively
invariant over the $z=0$ to 1 redshift range
(\markcite{lilly95}Lilly et al.\ 1995).
In order to make the most direct comparison,
we have chosen to construct an $R$-band LF
using our $z=0-1$ hard X-ray selected sample. In cases where
we do not have a measured $R$-band magnitude for a galaxy, 
we have estimated the $R$ magnitude using an average $R-I=0.6$ color 
(measured from galaxies in the CDF-N) and the galaxy's $I$ magnitude.
In computing the absolute rest-frame magnitudes, $M_R$, we have
used K-corrections appropriate for an Sb galaxy.

We first made a histogram of the absolute rest-frame 
magnitudes and then divided the numbers in the bins 
by the $z=0-1$ comoving volume.
Figure~\ref{fig8} shows the result and a comparison with the local 
$R$-band LFs of \markcite{lin96}Lin et al.\ (1996) and 
\markcite{geller97}Geller et al.\ (1997)
(the parameters for both were taken from the compilation by Geller et al.) 
The LF is X-ray flux selected and would increase
slowly as the flux selection was decreased below 
$1.0\times 10^{-15}$~erg~cm$^{-2}$~s$^{-1}$, but 
this effect may be expected to be small because
the cumulative counts converge rapidly below this flux limit.
Formally, the current X-ray selected LF is a lower bound to the 
luminosity density of galaxies showing X-ray activity.

If we integrate the local LFs and our histogram above $L^\ast$
(with $H_o=65$~km~s$^{-1}$~Mpc$^{-1}$, $M^\ast=-21.66$ for 
Geller et al.\ and $M^\ast=-21.58$ for Lin et al.), we find 
that $4\pm 1$\% of the galaxies are X-ray active.
If we extend the integral down to $0.5~L^\ast$, the percentage
is $1.5\pm 0.5$\% using Geller et al.\ and $2\pm 0.5$\% using
Lin et al. 

If the fraction of $>L^\ast$ galaxies showing such behavior 
reflects the fraction of time that each galaxy spends accreting onto its 
supermassive black hole, then each such galaxy must be active for 
about half a Gyr.
The duration of X-ray activity is much longer than the theoretically
estimated accretion time of 0.01~Gyr for black hole fuelling by a merger
(\markcite{kauffmann00}Kauffmann \& Haenelt 2000) and may suggest that
the accretion is being powered by many small mergers or by internal flows
within the galaxies, at least in the $z=0-1$ redshift range.

\section{Bolometric Luminosities}
\label{seclum}

Following B01, we use our multiwavelength dataset to estimate bolometric
luminosities ($L_{BOL}=L_{FIR}+L_{OPT}+L_{X}$) for our combined hard
X-ray selected sample. For the spectroscopically unidentified sources,
we nominally assume $z=2$.

As in \S~3.3 of B01, we allow for the average effects of
opacity in the calculation of our hard X-ray luminosities
by normalizing the flux at 4~keV for a $\Gamma=2$ spectrum to the
flux at 4~keV calculated over the $2-7$~keV energy range for
a spectrum with photon index $\Gamma=1.2$. Then

\begin{equation}
L_{HX}=4~\pi~d_L^2~f_{HX}
\end{equation}

\noindent
where $f_{HX}$ is the $2-7$~keV flux of Table~\ref{tab1} and
Table~\ref{tab3}. For this energy range the $L_{HX}$ equation would
be the same if we had instead used $\Gamma=1.4$.
The inferred hard X-ray luminosities for the sources in the
A370 and CDF-N fields are given in column~3 of Table~\ref{tab6}.

The total X-ray luminosity is only weakly sensitive to
the adopted energy range for a photon index $\Gamma=2$.
For an energy range from 0.1 to 100~keV,
the ratio of the total X-ray luminosity, $L_X$, to the $2-7$~keV
luminosity is $\ln(1000)/\ln(3.5)=5.5$.
The total X-ray luminosities of the sources are 
smaller than the contributions to the bolometric luminosities
from other wavelengths.

We estimate the luminosities in the ultraviolet/optical using the
equation given in \S~4.3 of B01, 

\begin{equation}
L_{OPT}=4~\pi~d_L^2~(9.4\times 10^{15})~f_{2500(1+z)}~(1+z)^{-1}
\end{equation}

\noindent
where $d_L$ is the luminosity distance in cm and $f_{2500(1+z)}$
has units erg~cm$^{-2}$~s$^{-1}$~Hz$^{-1}$. For the CDF-N sample,
$f_{2500(1+z)}$ is
interpolated from the observed fluxes that correspond to the magnitudes
given in Table~\ref{tab3}, and, where available, the measured
$R$ magnitudes from H01 and the $U'$ (3400\AA) magnitudes from
G. Wilson, et al., in preparation. 
For sources that have substantial galaxy light contamination, 
$L_{OPT}$ is an upper limit on the AGN contribution in
the ultraviolet/optical.

For the A370 field, where we have substantially less imaging data,
we have only been able to calculate $L_{OPT}$ for
the five sources in Table~\ref{tab1}
with sufficient magnitude coverage --- given the measured or assumed
redshifts --- to allow reliable interpolations (sources 3, 4, 5, 6, and 12).
The inferred bolometric ultraviolet/optical luminosities for the 
A370 and CDF-N sources are listed in column~4 of Table~\ref{tab6}.

The bolometric far-infrared (FIR) flux is related to the rest-frame 
20~cm flux through the well-established FIR-radio correlation
(\markcite{condon92}Condon 1992) of local starburst galaxies and
radio-quiet AGN. The correlation also seems to hold at high redshift 
(e.g., \markcite{cy00}Carilli \& Yun 2000; 
\markcite{bcr00}Barger, Cowie, \& Richards 2000).
It is not entirely known what mechanism produces the relationship. 
In galaxies without a powerful AGN, the radio luminosity is dominated
by diffuse synchrotron emission;
thus, the tight empirical correlation is thought to be a consequence of 
the radio continuum emission and the thermal dust emission
both being linearly related to the massive star formation rate.
In the present systems a substantial fraction of the 
FIR luminosity may also be of AGN origin rather than from star formation,
but the only strong
statement that we can make is that the FIR luminosity from the AGN 
is bounded above by the FIR-radio correlation result from \S 6.3 of B01,

\begin{equation}
L_{FIR}=4~\pi~d_L^2~(8.4\times 10^{14})~f_{20}~(1+z)^{-0.2}
\end{equation}

\noindent
Here $d_L$ is in cm and
$f_{20}$ has units erg~cm$^{-2}$~s$^{-1}$~Hz$^{-1}$.

The inferred bolometric FIR luminosities or limits
for the hard X-ray sources in the A370 and CDF-N fields
are given in column~5 of Table~\ref{tab6}.
The luminosities for the sources that are not detected
above the radio $3\sigma$ limit are calculated as upper 
limits based on the $3\sigma$ limit.
The SSA13 hard X-ray, optical, and FIR luminosities can be found
in Table~2 of B01.

In Fig.~\ref{fig9} we plot the ratios
$L_{FIR}/L_{HX}$ (open symbols) and $L_{OPT}/L_{HX}$ (filled symbols)
versus redshift for the spectroscopically identified sources
for which we were able to estimate $L_{OPT}$. 
The FIR light dominates over the optical light in a high fraction (73\%) 
of the sources. For the sources in the A370 field for which we
were unable to estimate $L_{OPT}$, we take 
$L_{FIR}$ to be the upper limit to the bolometric luminosity.
For all sources we are neglecting possible contributions in the
extreme ultraviolet, for which we do not have data.

%
%

\begin{inlinefigure}
\psfig{figure=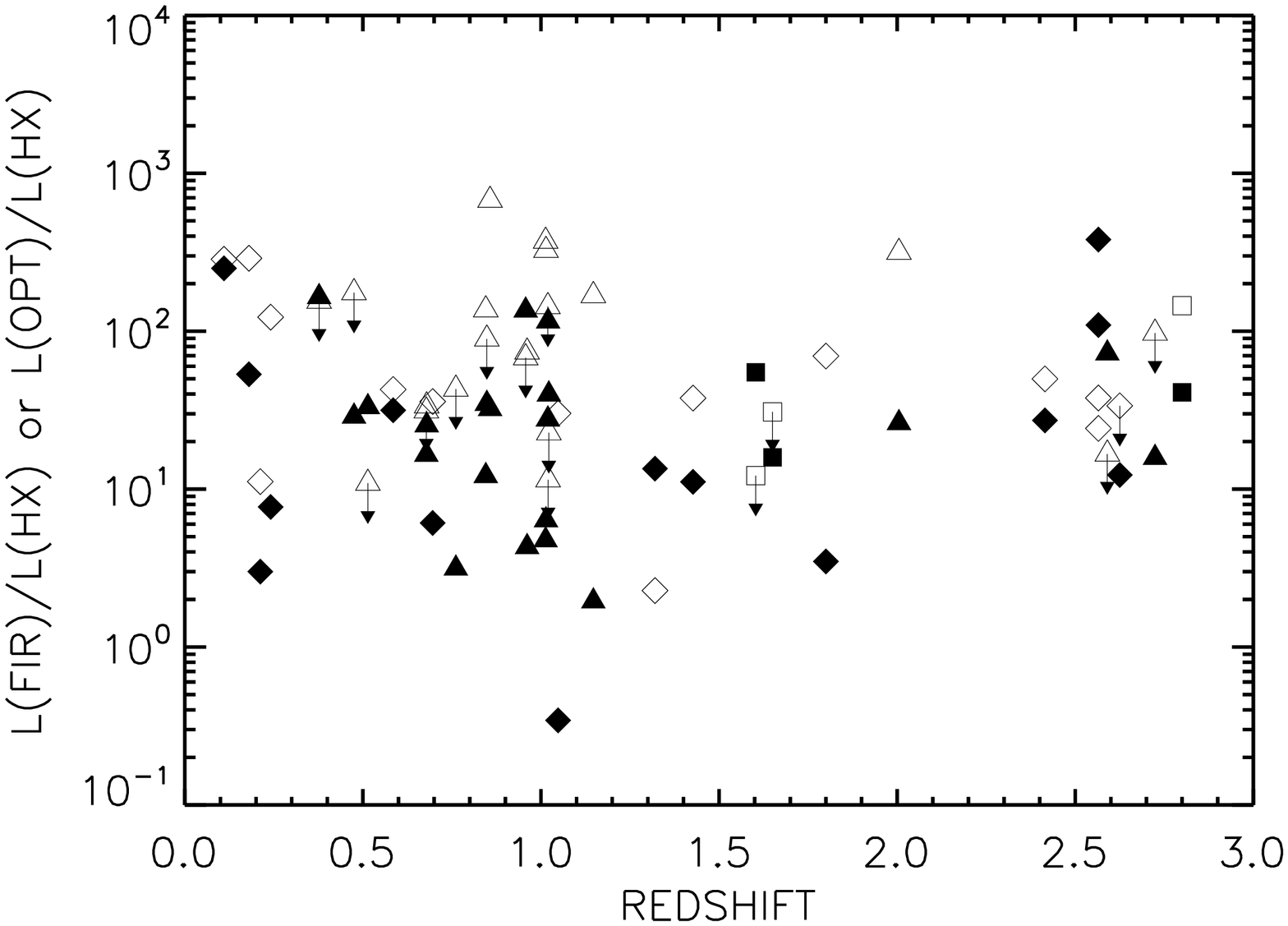,angle=90,width=3.5in}
\vspace{6pt}
\figurenum{9}
\caption{
Ratio of the bolometric optical (filled) and radio inferred FIR (open)
luminosities to the hard X-ray luminosities versus redshift for the
A370 (squares), CDF-N (triangles), and SSA13 (diamonds) fields.
Spectroscopically unidentified sources are not shown.
Downward pointing arrows indicate the use of $3\sigma$ radio limits 
in estimating the FIR luminosities.
}
\label{fig9}
\addtolength{\baselineskip}{10pt}
\end{inlinefigure}

\section{Accretion History}

The mass inflow rate into a black hole, $dM/dt$ or $\dot{M}_{BH}$, 
is related to the AGN's bolometric 
luminosity by $\epsilon\dot{M}_{BH} = L_{BOL}/c^2$, where $\epsilon$ 
is the radiative efficiency of the accretion energy.
With $\dot{M}_{BH}$ in units of $M_\odot$~Gyr$^{-1}$, the relation is

\begin{equation}
\dot{M}_{BH}=1.76\times 
10^5~(0.1/\epsilon)~(L_{BOL}/10^{42})
\end{equation}

\noindent
where $L_{BOL}$ is in units of erg~s$^{-1}$.

If we make the simplifying assumption that $\epsilon$ has an 
approximately universal value, $\epsilon\sim 0.1$, we can evaluate
$\dot{M}_{BH}$ for our hard X-ray sources. 
We calculate $\dot{M}_{BH}$ upper bounds from our estimated 
values of $L_{BOL}$ (\S\ref{seclum}) and lower bounds with 
$L_{BOL}$ replaced by $L_{X}$.

%
%
 
\begin{inlinefigure}
\psfig{figure=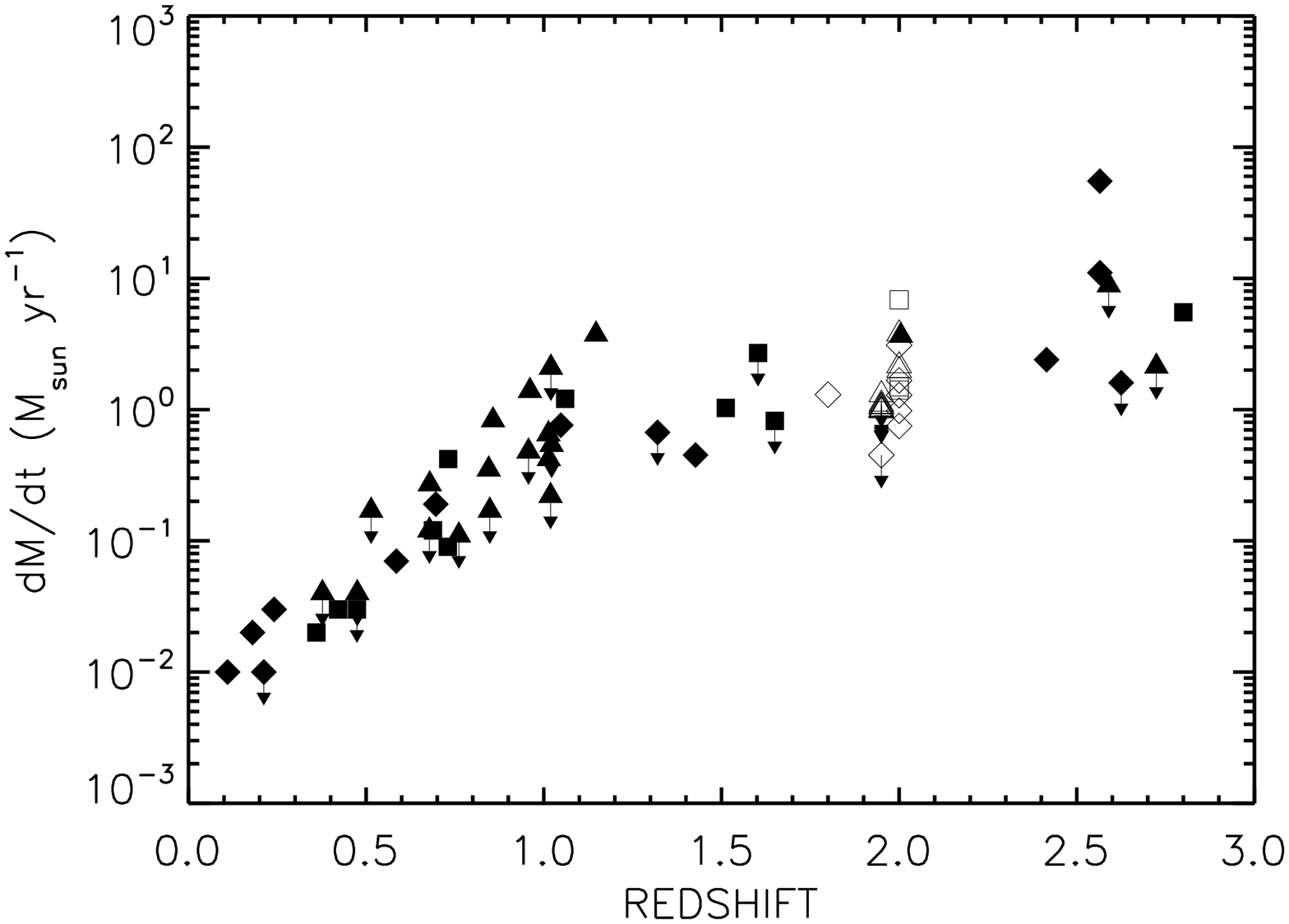,angle=90,width=3.5in}
\psfig{figure=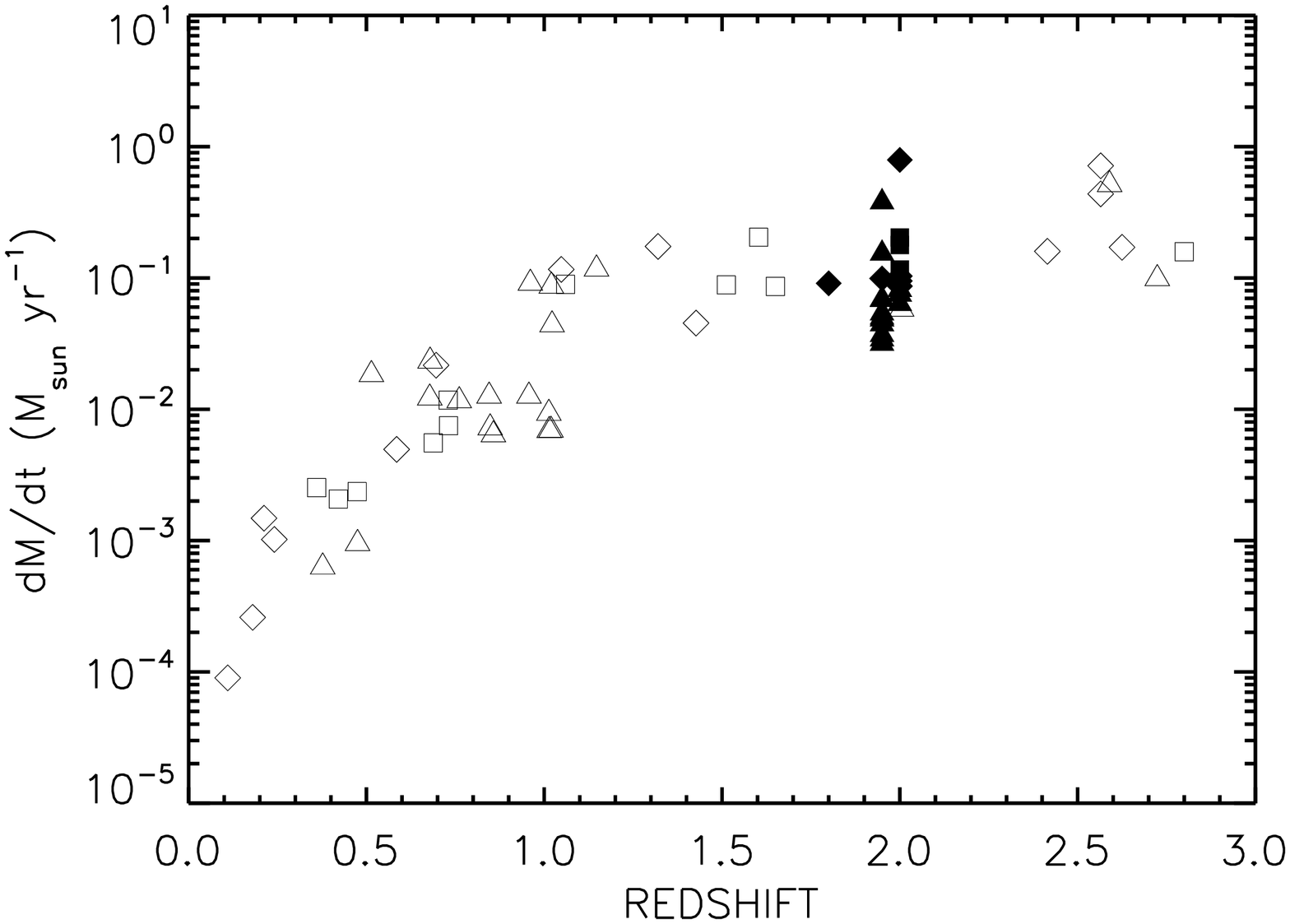,angle=90,width=3.5in}
\vspace{6pt}
\figurenum{10}
\caption{
The mass inflow rate ($dM_{BH}/dt$ or $\dot{M}_{BH}$)
versus redshift for the hard X-ray sources in
the A370 (squares), CDF-N (triangles), and SSA13 (diamonds)
fields calculated using either (a)~the bolometric luminosities, $L_{BOL}$,
or (b)~the bolometric X-ray luminosities, $L_X$.
Downward pointing arrows in (a) indicate the use of $3\sigma$
radio limits in calculating the FIR luminosities.
In (a), spectroscopically identified sources are denoted by filled 
symbols and spectroscopically unidentified sources are denoted by 
open symbols, nominally placed at $z=2$ (a slight offset in redshift has
been applied to the unidentified sources with radio limits). 
In (b), we invert the above for contrast, such that the spectroscopically 
identified sources are denoted by open symbols, and the spectroscopically 
unidentified sources are denoted by filled symbols.
The source in the SSA13 field with a millimetric redshift of 1.8 
(as deduced from the submillimeter to radio flux ratio; see B01) 
is shown as a diamond at $z=1.8$.
}
\label{fig10}
\addtolength{\baselineskip}{10pt}
\end{inlinefigure}

Figure~\ref{fig10} shows the $\dot{M}_{BH}$ values versus redshift
for the canonical $\epsilon=0.1$. The upper bounds in Fig.~\ref{fig10}a 
are on average about a factor 20 higher than the lower bounds 
in Fig.~\ref{fig10}b, but the trends with redshift discussed below 
are similar for both.
To generate a $10^9$~M$_\odot$ black hole over an
accretion period of order 0.5~Gyr, an $\dot{M}$ of the order of
$2~M_\odot$~yr$^{-1}$ is required.

The lower envelope of the $\dot{M}_{BH}$ versus redshift plot simply
reflects selection bias, since high redshift sources with low $\dot{M}_{BH}$ 
will not be detected in the current hard X-ray surveys.
The upper envelope of the distribution strongly reflects the changing
strength of the maximal accretion rate with redshift, although
some of the change is due to the volume increase
with redshift, which gives a higher chance of finding
very luminous sources ($\Delta V$ for a $z=2-3$ bin
is a factor $3.3$ larger than for a $z=0-1$ bin).
At $z\gtrsim1$ the inferred accretion rates are up to
two orders of magnitude larger than those at $z\lesssim 0.5$.
Thus, the most violent episodes of black hole growth occurred at
early times. For $z\lesssim 1$, $\log(\dot{M}_{BH})$
increases approximately linearly with redshift.
If the {\it Chandra} surveys are still missing AGN that suffer from Compton
thick absorption, then these sources would selectively appear at lower redshifts
since the effective column density is a factor of $(1+z)^{2.6}$ lower
than the true column density (B01).

%
%

\begin{inlinefigure}
\psfig{figure=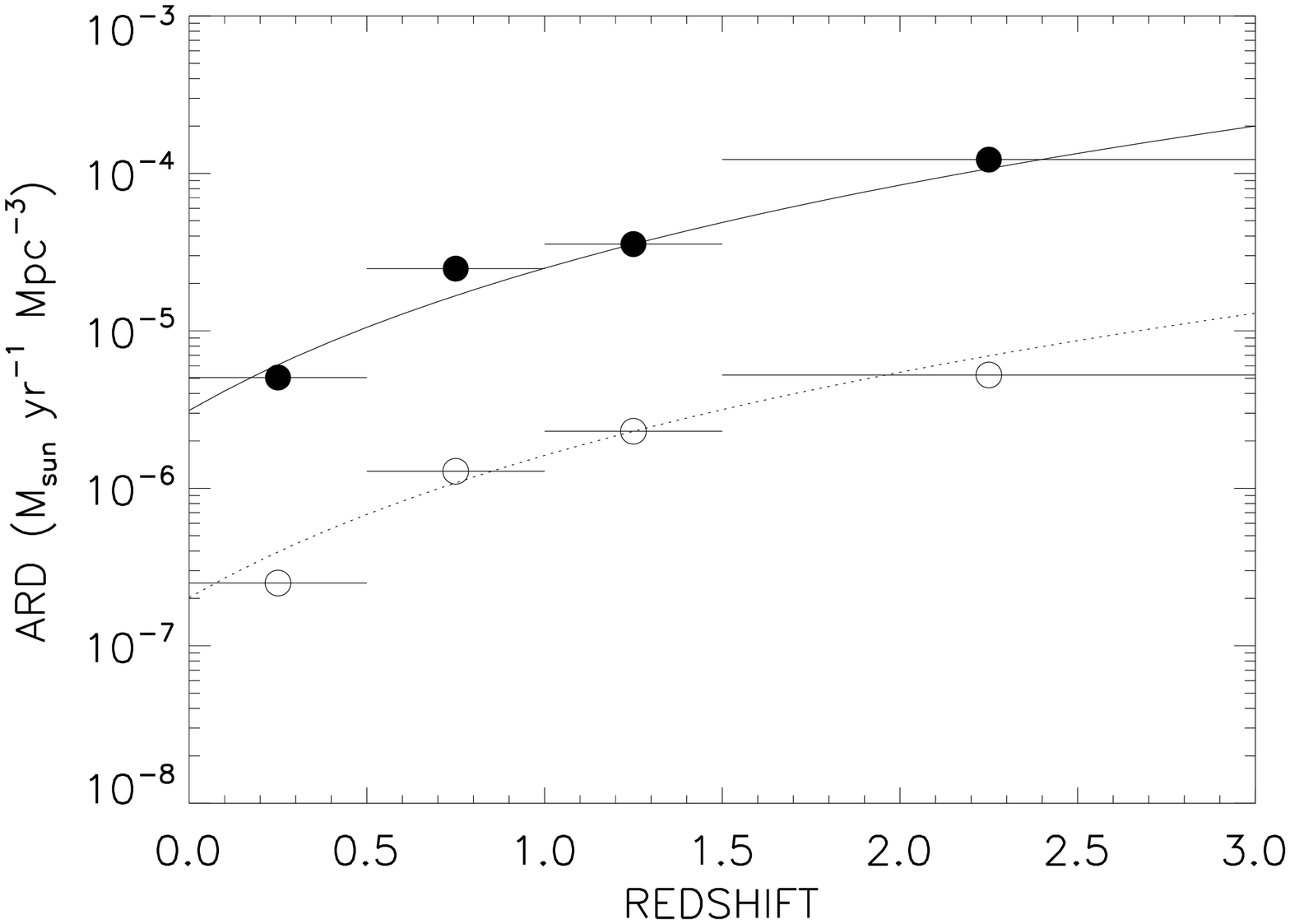,angle=90,width=3.5in}
\vspace{6pt}
\figurenum{11}
\caption{
The accretion rate density ($d\rho_{BH}/dt$ or $\dot{\rho}_{BH}$)
of the combined hard X-ray sample versus
redshift for four redshift bins: $0<z<0.5$, $0.5\le z<1$,
$1\le z<1.5$, and $1.5\le z\le 3$. The solid data points are
calculated from the bolometric luminosities, and the open data 
points are calculated from the bolometric X-ray luminosities. 
The curves illustrate a $(1+z)^3$ dependence, normalized to the
$1\le z<1.5$ redshift bin.
}
\label{fig11}
\addtolength{\baselineskip}{10pt}
\end{inlinefigure}

From the $\dot{M}_{BH}$ versus redshift distribution, we can deduce
the time history of the accretion rate density (ARD),
$\dot{\rho}_{BH}=\sum \dot{M}_{BH}/\Delta V$, where the sum is over
the sources in a bin $\Delta z$ and $\Delta V$ is the comoving volume
element in Mpc$^3$, which is proportional to the area coverage.
The A370, CDF-N, and SSA13 fields have areas of
56, 78, and 64~arcmin$^2$, respectively, giving a total area
$A=198$~arcmin$^2$.
The maximal and minimal results for the ARD history, determined from
the upper and lower bounds of Fig.~\ref{fig10},
are given in Fig.~\ref{fig11}. Again, this
is  normalized with a value of $\epsilon\sim 0.1$, and the accretion
rates scale as $\epsilon^{-1}$.

The time history of the ARD seems to fall roughly as $(1+z)^3$, 
as illustrated by the overlaid curves in Fig.~\ref{fig11}, 
although the evolution
could be slower considering the present limitations in deducing the
mass inflow rates. This redshift dependence is similar to
that inferred in star formation histories (or, more directly,
in rest-frame ultraviolet light densities) where slopes over the
$z=0-1$ interval have measured values in the range 1.5 to 4
(\markcite{lilly96}Lilly et al.\ 1996;
\markcite{cowie99}Cowie, Songaila, \& Barger 1999;
G. Wilson, et al., in preparation).

We integrate the maximal accretion rate density from $z=0$ to $z=3$ 
to obtain a current black
hole mass density, $\rho_{BH}=6\times 10^{-35}$~g~cm$^{-3}$.
This density can be compared to the spheroidal mass density of 
$10^{-32}$~g~cm$^{-3}$, which is accurate to about a multiplicative 
factor of 2 (\markcite{cowie88}Cowie 1988; 
\markcite{fukugita98}Fukugita et al.\ 1998).
The ratio $\rho_{BH}/\rho_{SPH}=0.006$
is reasonably consistent with the black hole-bulge mass relation
$M_{BH}=0.001$---0.002~$M_{bulge}$ found locally
(e.g., \markcite{wandel99}Wandel 1999;
\markcite{ferr00}Ferrarese \& Merritt 2000;
\markcite{gebhardt00}Gebhardt et al.\ 2000).

\section{Summary}

The bulk of the hard X-ray background has already been resolved by
deep {\it Chandra} surveys. In this paper we have presented follow-up 
multiwavelength studies of the AGN sources that comprise the background
to ascertain their nature.
By combining datasets from our previous studies of the CDF-N and SSA13 
fields with new results on the A370 cluster field, we have amassed a 
large sample of 69 hard X-ray sources, of which 45 now have redshifts.
With our optical, near-infrared, and 20~cm images of the fields, we have
made a general analysis of AGN accretion phenomena. Our principal
findings are as follows: 

(i)~About 4\% of the $>L^\ast$ galaxy population is
X-ray luminous at any time, and hence the average duration of supermassive
black hole accretion is about half a Gyr. Accretion activity might 
occur over this long a period in all galaxies, or some galaxies might be 
active for much longer periods while others are completely inactive.

(ii)~We determined the X-ray, optical,
and FIR luminosities of the hard X-ray sources. 
We then estimated the mass inflow rates onto the supermassive 
black holes, assuming a canonical radiative efficiency $\epsilon=0.1$.  
We determined maximal inflow rates using 
$L_{BOL}=L_{FIR}+L_{OPT}+L_{X}$ (this assumes the dominant component,
$L_{FIR}$, is entirely due to reradiation of AGN light) 
and minimal rates using only the bolometric X-ray luminosities.
These estimates differ by about a multiplicative factor of 20.
Over the range $z=0.1$ to $z=1$, $\log \dot{M}_{BH}$ 
increases approximately linearly with redshift.
Above $z=1$, $\log\dot{M}_{BH}$ flattens with redshift
with a substantial scatter that may be due to a range of efficiencies,
$\epsilon$, or to differences in the luminosities;
the absence of sources at $z>1$ with $\log\dot{M}_{BH}<0.1$
is likely a selection effect.
The maximal mass inflow rates increase from $\sim 0.01$~M$_\odot$~yr$^{-1}$
at $z\lesssim 0.5$ up to $\sim10$~M$_\odot$~yr$^{-1}$ at $z\gtrsim 1$.

(iii)~We estimated the accretion rate density, $d\rho_{BH}/dt$,
versus redshift. The time history of the accretion rate density 
seems to fall roughly as $(1+z)^3$. This is a similar dependence
to that inferred in star formation histories.
For the maximal ARD, the integrated value 
is $\rho_{BH}=6\times 10^{-35}$~g~cm$^{-3}$,
which is reasonably consistent with the local black hole mass
relation combined with the estimated spheroidal density. 

\acknowledgements
AJB acknowledges support from NASA through Hubble
Fellowship grant HF-01117.01-A awarded by the
Space Telescope Science Institute, which is operated by the
Association of Universities for Research in Astronomy, Inc.,
for NASA under contract NAS 5-26555.
We gratefully acknowledge support from NSF grants
AST-0084847 (AJB) and AST-0084816 (LLC), 
NSF CAREER award AST-9983783 (WNB),
NASA contracts NAS 8-37716 (MWB) and NAS 8-38252 (GPG, PI),
NASA GSRP grant NGT 5-50247 and the Pennsylvania Space
Grant Consortium (AEH).

\newpage

\begin{deluxetable}{rrrccccccccr}
\renewcommand\baselinestretch{1.0}
\tablewidth{0pt}
\parskip=0.2cm
\tablenum{1}
\footnotesize
\tablecaption{Hard X-ray Sources in the A370 Field}
\tablehead{
\# & RA(2000) & Dec(2000) & $f(2-7\ {\rm keV})$ & $f(0.5-2\ {\rm keV})$
& $HK'$ & $I$ & $R$ & $V$ & $z$ & $\Delta\theta$ & $S(20 {\rm cm})$ \cr
& & & $(10^{-15}$\ erg\ cm$^{-2}$\ s$^{-1})$ &
$(10^{-16}$\ erg\ cm$^{-2}$\ s$^{-1})$ & & & & & & (arcsec) & ($\mu$Jy)
}
\startdata
0 & 2 40 06.72 & $-1$ 36 56.2 & $2.9\pm 0.9$ & $22.8\pm 3.3$ 
& \nodata & 18.6 & 19.8 & \nodata & 0.421\tablenotemark{n} & 0.6 & 30 \cr

1 & 2 40 04.76 & $-1$ 38 11.6 & $5.7\pm 1.5$ & \nodata 
& \nodata & 23.4 & \nodata & \nodata & \nodata & 1.4 & 182 \cr

2 & 2 40 00.86 & $-1$ 33 13.2 & $4.4\pm 1.2$ & $10.5\pm 2.3$ 
& 17.6 & 20.1 & 21.5 & 22.3 & 0.730\tablenotemark{n} & 0.5 & 25 \cr

3 & 2 40 00.25 & $-1$ 32 34.1 & $4.4\pm 1.2$ & $22.6\pm 3.3$ 
& \nodata & 22.1 & 23.0 & 23.6 & 1.650\tablenotemark{b} & 0.3 & $<20$ \cr

4 & 2 39 58.98 & $-1$ 35 48.9 & $11.2\pm 1.8$ & $45.3\pm 4.7$ 
& 18.3 & 20.0 & 20.6 & 21.0 & 1.603\tablenotemark{b} & 0.3 & $<20$ \cr

5 & 2 39 58.62 & $-1$ 32 59.9 & $6.5\pm 1.4$ & $26.1\pm 3.6$ 
& 18.8 & 23.0 & 24.2 & 24.8 & \nodata & 0.4 & 29 \cr

6 & 2 39 58.14 & $-1$ 32 36.1 & $3.7\pm 1.1$ & \nodata 
& \nodata & 23.0 & 24.1 & 24.8 & \nodata & 0.3 & 29 \cr

7 & 2 39 57.92 & $-1$ 38 52.5 & $2.8\pm 1.0$\tablenotemark{a} & \nodata 
& \nodata & 21.7 & \nodata & \nodata & 0.731\tablenotemark{n} & 0.8 & 116 \cr

8 & 2 39 57.83 & $-1$ 37 05.9 & $2.4\pm 0.9$ & $13.5\pm 2.6$ 
& \nodata & 23.1 & \nodata & \nodata & 0.688\tablenotemark{n} & 0.4 & 39 \cr

9 & 2 39 57.74 & $-1$ 36 04.6 & $2.5\pm 0.9$ & \nodata 
& 18.4 & 20.8 & 21.8 & 22.8 & 0.474\tablenotemark{h} & 0.4 & $<20$ \cr

10 & 2 39 57.17 & $-1$ 32 59.4 & $5.1\pm 1.3$ & \nodata 
& 15.8 & 18.3 & 19.3 & 20.4 & 0.360\tablenotemark{n} & 0.2 & 28 \cr

11\tablenotemark{c} & 2 39 56.59 & $-1$ 34 26.6 & $28.3\pm 2.9$ & $9.5\pm 2.5$
& 16.8 & 19.6 & 20.9 & 21.7 & 1.060\tablenotemark{h} & 0.3 & 296 \cr

12\tablenotemark{d} & 2 39 51.83 & $-1$ 35 58.5 & $5.4\pm 1.3$ & \nodata 
& 18.4 & 21.2 & 21.9 & 22.6 & 2.800\tablenotemark{h} & 0.4 & 124 \cr

13 & 2 39 49.66 & $-1$ 38 45.4 & $5.6\pm 1.4$ & $89.0\pm 6.7$ 
& \nodata & 20.9 & \nodata & \nodata & 1.512\tablenotemark{b} & 0.5 & 52 \cr

14 & 2 39 49.22 & $-1$ 36 57.3 & $13.1\pm 2.0$ & $65.5\pm 5.7$ 
& \nodata & 21.3 & 21.8 & 22.2 & 3.268\tablenotemark{b} & 0.4 & 28
\enddata
\label{tab1}
\tablecomments{
$^{\rm a}$Flux based on counts measured in a $2''$ radius aperture.
$^{\rm c}$Flux and magnitudes uncorrected for magnification by a factor of 2.1.
$^{\rm d}$Flux and magnitudes uncorrected for magnification by a factor of 2.4;
spectrum from Ivison et al.\ (1998).
In column (10) the superscript ``b'' denotes sources with broad emission lines,
``h'' denotes sources with high ionization lines, and
``n'' denotes a ``normal'' galaxy spectrum.
Source 13 coordinates are the means of the positions in the soft and hard
bands. For source 11, which appears to be a radio jet, we have used the flux
associated with a bright peak near the X-ray source, which may be considered
an upper limit on the flux associated with the object. For source 12 we
have used the radio flux at the X-ray position, excluding emission which appears
to arise in the neighboring optical emission-line object. Source 1 appears
to be a radio double, and the radio flux may not be directly associated with
the X-ray source.
}
\end{deluxetable}

\begin{deluxetable}{rrrccr}
\renewcommand\baselinestretch{1.0}
\tablewidth{0pt}
\parskip=0.2cm
\tablecaption{Supplemental Soft X-ray Sources in the A370 Field}
\tablenum{2}
\small
\tablehead{
\# & RA(2000) & Dec(2000) & $f(0.5-2\ {\rm keV})$
& $z$ & $S(20 {\rm cm})$ \cr
& & & $(10^{-16}$\ erg\ cm$^{-2}$\ s$^{-1})$ & & ($\mu$Jy)
}
\startdata
S0 & 2 40 10.60 & $-1$ 37 04.7 & $3.1\pm 1.2$ 
& 3.025 & $<20$ \cr

S1 & 2 40 06.17 & $-1$ 34 01.0 & $6.7\pm 1.8$ 
& 0.553 & 40 \cr

S2 & 2 40 05.30 & $-1$ 36 31.2 & $8.8\pm 2.1$ 
& \nodata & $<20$ \cr

S3 & 2 40 03.70 & $-1$ 39 43.1 & $4.1\pm 1.5$ 
& \nodata & $<20$ \cr

S4 & 2 40 02.16 & $-1$ 35 09.0 & $5.1\pm 1.6$ 
& \nodata & 37 \cr

S5 & 2 40 00.92 & $-1$ 35 16.2 & $6.2\pm 1.8$ 
& 0.380 & $<20$ \cr

S6 & 2 40 00.12 & $-1$ 31 58.1 & $7.0\pm 1.9$ 
& \nodata & $<20$ \cr

S7 & 2 39 58.74 & $-1$ 37 50.4 & $5.2\pm 1.6$ 
& \nodata & $<20$ \cr

S8 & 2 39 58.64 & $-1$ 33 09.2 & $22.5\pm 3.3$ 
& 1.303 & 21 \cr

S9 & 2 39 56.35 & $-1$ 31 36.9 & $28.5\pm 3.8$ 
& 0.026 & 883 \cr

S10 & 2 39 54.64 & $-1$ 32 33.5 & $199.\pm 10.$ 
& 0.045 & 358 \cr

S11 & 2 39 52.93 & $-1$ 34 26.2 & $8.6\pm 3.1$ 
& \nodata & $<20$ \cr

S12 & 2 39 47.53 & $-1$ 35 12.2 & $7.7\pm 2.1$ 
& 1.605 & $<20$ \cr

S13 & 2 39 40.30 & $-1$ 34 31.2 & $11.6\pm 2.5$ 
& 0.423 & 76 \cr

\enddata
\label{tab2}
\end{deluxetable}

\newpage

\begin{deluxetable}{rcrrrrrrrrcc}
\renewcommand\baselinestretch{1.0}
\tablewidth{0pt}
\parskip=0.2cm
\tablenum{3}
\footnotesize
\tablecaption{Hard X-ray Sources in the CDF-N Field}
\tablehead{
\# & CXOHDFN Name & $f(2-7~{\rm keV})$ & $f(0.5-2~{\rm keV})$
& $HK'$ & $I$ & $V$ & $B$ & $z$ & $S(20 {\rm cm})$ \cr
& (J2000) & \multicolumn{2}{c}{$(10^{-15}$~erg~cm$^{-2}$~s$^{-1})$} 
& & & & & & ($\mu$Jy)
}
\startdata
 0 & 123615.9+621515 &  2.39 &  0.19 &
$>21.5$ & 24.4 & 26.1 & $>26.1$ & \nodata & 54 \cr

 1 & 123616.0+621107 & 12.10 &  1.05 &
21.1 & 24.4 & $>26.4$  & $>26.1$ & \nodata & $<23$ \cr

 2 & 123617.0+621011 &  3.31 &  1.87 &
18.7 & 21.7 & 24.0 & 24.3 & 0.845\tablenotemark{n} & 62 \cr

 3 & 123618.0+621635 & 10.40 &  6.32 &
18.0 & 19.9 & 21.4 & 22.3 & 0.679\tablenotemark{n} & 47 \cr

 4 & 123618.5+621115 &  7.25 &  7.66 &
18.2 & 21.0 & 22.0 & 21.8 & 1.022\tablenotemark{b} & $<23$ \cr

 5 & 123619.1+621441 &  1.42 &  1.34 &
21.1 & 24.3 & 25.2 & $>26.1$ & \nodata & $<23$ \cr

 6 & 123621.3+621109 &  1.16 &  \nodata &
18.3 & 21.9 & 26.1 & 25.6 & 1.014\tablenotemark{n} & 53 \cr

 7 & 123622.6+621028 &  2.18 &  1.09 &
20.9 & 24.7 & 25.9 & $>26.1$ & \nodata & $<23$ \cr

 8 & 123622.9+621527 &  8.81 &  7.69 &
18.7 & 20.0 & 20.5 & 20.4 & 2.590\tablenotemark{b} & $<23$ \cr

 9 & 123627.5+621026 &  3.94 &  0.25 &
20.4 & 22.8 & 25.2 & 25.9 & 0.761\tablenotemark{n} & $<23$ \cr

10 & 123627.7+621158 &  1.09 &  \nodata &
$>21.5$ & 23.7 & 25.7 & $>26.1$ & \nodata & $<23$ \cr

11 & 123629.0+621046 &  1.57 &  \nodata &
18.9 & 22.8 & 25.8 & $>26.1$ & 1.013\tablenotemark{n} & 81 \cr

12 & 123629.2+621613 &  1.88 &  \nodata &
18.3 & 20.4 & 23.4 & 23.8 & 0.848\tablenotemark{n} & $<23$ \cr

13 & 123635.6+621424 & 1.84 &  0.36 &
19.4 & 22.9 & 23.8 & 24.1 & 2.005\tablenotemark{h} & 88 \cr

14 & 123636.6+621347 &  2.44 &  2.94 &
19.0 & 21.1 & 22.1 & 22.6 & 0.957\tablenotemark{h} & $<23$ \cr

15 & 123642.2+621711 &  1.50 &  0.97 &
21.4 & 23.7 & 25.2 & 25.8 & 2.724\tablenotemark{b} & $<23$ \cr

16 & 123642.2+621545 &  1.62 &  0.41 &
18.1 & 21.2 & 23.3 & 24.2 & 0.857\tablenotemark{h} & 150 \cr

17 & 123646.3+621404 & 17.48 &  2.82 &
18.2 & 21.0 & 22.8 & 23.7 & 0.961\tablenotemark{h}  & 179 \cr

18 & 123646.6+620857 &  1.01 &  \nodata &
19.2 & 23.0 & $>26.4$ & $>26.1$ & \nodata & $<23$ \cr

19 & 123651.2+621051 &  1.72 &  \nodata &
19.6 & 23.3 & $>26.4$ & $>26.1$ & \nodata & $<23$ \cr

20\tablenotemark{a} & 123651.7+621221 &  2.60 &  0.18 &
$>22.6$ & $>25.3$ & 22.9 & 23.6 & \nodata & 49 \cr

21 & 123658.8+621435 &  5.47 &  2.48 &
17.9 & 20.4 & 22.5 & 23.0 & 0.678\tablenotemark{n}  & $<23$ \cr

22 & 123702.7+621543 & 16.03 &  5.41 &
17.2 & 19.2 & 21.0 & 21.0 & 0.514\tablenotemark{n}  & $<23$ \cr

23 & 123704.6+621652 &  1.15 &  1.18 &
18.2 & 20.1 & 21.4 & 22.3 & 0.377\tablenotemark{h}  & $<23$ \cr

24 & 123704.8+621601 &  4.91 &  2.91 &
$>21.5$ & 25.2 & $>26.4$ & $>26.1$ & \nodata & $<23$ \cr

25 & 123706.8+621702 & 14.38 & 14.72 &
18.0 & 19.6 & 20.0 & 20.1 & 1.020\tablenotemark{b}  & $<23$ \cr

26 & 123713.7+621424 &  1.00 &  \nodata &
18.5 & 21.2 & 23.2 & 24.1 & 0.475\tablenotemark{n}  & $<23$ \cr

27 & 123714.1+620916 &  1.60 &  0.84 &
20.2 & 24.0 & 26.0 & 26.0 & \nodata & $<23$ \cr

28 & 123714.8+621617 &  1.18 &  0.66 &
20.4 & 23.3 & 24.0 & 24.3 & \nodata & $<23$ \cr

29 & 123715.9+621213 &  1.15 &  \nodata &
19.2 & 22.3 & 25.0 & 25.5 & 1.019\tablenotemark{h} & $<23$ \cr

30 & 123716.7+621733 & 14.56 &  4.57 &
18.8 & 22.2 & 24.3 & 24.9 & 1.147\tablenotemark{h}  & 346 \cr

31 & 123724.0+621304 &  2.20 &  0.27 &
20.4 & 23.6 & 25.2 & 25.3 & \nodata & $<23$ \cr

32 & 123724.3+621359 &  1.55 &  \nodata &
20.9 & 24.1 & 25.8 & 25.5 & \nodata & $<23$ \cr

33 & 123725.0+620856 &  2.03 &  0.41 &
18.7 & 22.2 & 25.6 & $>26.1$ & \nodata &  90
\enddata
\label{tab3}
\tablecomments{
$^{\rm a}$X-ray coordinate offset from bright galaxy at 
$z=0.401$ is $1.2''$; radio coordinate offset from
bright galaxy is $1.4''$.
In column (10) the superscript ``b'' denotes sources with broad emission lines,
``h'' denotes sources with high ionization lines, and 
``n'' denotes a ``normal'' galaxy spectrum.
}
\end{deluxetable}

\begin{deluxetable}{ccccc}
\renewcommand\baselinestretch{1.0}
\tablewidth{0pt}
\tablecaption{Optical to Hard X-ray Fluxes}
\parskip=0.2cm
\tablenum{4}
\small
\tablehead{Range & \# Sources & $\langle f(2-7~{\rm keV})\rangle$ &
$\langle I \rangle$ & median $I$ \cr
($10^{-15}$~erg~cm$^{-2}$~s$^{-1}$) & & ($10^{-15}$~erg~cm$^{-2}$~s$^{-1}$)}
\startdata
1--2 & 19 & 1.4 & 21.8 & 23.0 \cr
2--5 & 27 & 3.0 & 19.0 & 22.8 \cr
$>5$ & 23 & 11.3 & 19.8 & 20.5
\enddata
\label{tab4}
\end{deluxetable}

\begin{deluxetable}{ccccc}
\renewcommand\baselinestretch{1.0}
\tablewidth{0pt}
\tablecaption{Spectral Classes for the A370, CDF-N, and SSA13 Hard X-ray Samples}
\parskip=0.2cm
\tablenum{5}
\small
\tablehead{Class & A370 & CDF-N & SSA13 & Summed}
\startdata
Broad Line & 4 & 4 & 2 & 10 \cr
High Ionization & 3 & 7 & 5 & 15 \cr
`Normal' & 5 & 9 & 6 & 20 \cr
Unidentified $I\le 23.5$ & 3 & 4 & 0 & 7 \cr
Unidentified $I>23.5$ & 0 & 10 & 7 & 17\cr
\enddata
\label{tab5}
\end{deluxetable}

\begin{deluxetable}{ccrrr}
\renewcommand\baselinestretch{1.0}
\tablewidth{0pt}
\tablecaption{Luminosities for the A370 and CDF-N Hard X-ray Samples}
\parskip=0.2cm
\tablenum{6}
\small
\tablehead{Number & $z$ & $L_{HX}$ & $L_{OPT}$ & $L_{FIR}$ \cr 
& & ($10^{42}$~ergs~s$^{-1}$) & ($10^{42}$~ergs~s$^{-1}$) & 
($10^{42}$~ergs~s$^{-1}$)}
\startdata
\sidehead{A370}
 0 & 0.421 &      2.1 & \nodata &   170    \cr
 1 & 2.000 &      180 & \nodata & 39000    \cr
 2 & 0.730 &       12 & \nodata &   510    \cr
 3 & 1.650 &       89 &    1400 &  $<2700$ \cr
 4 & 1.603 &      210 &   12000 &  $<2600$ \cr
 5 & 2.000 &      210 &    1100 &  6200    \cr
 6 & 2.000 &      120 &    1200 &  6200    \cr
 7 & 0.731 &      7.8 & \nodata &  2400    \cr
 8 & 0.688 &      5.7 & \nodata &   690    \cr
 9 & 0.474 &      2.4 & \nodata &  $<150$  \cr
10 & 0.360 &      2.6 & \nodata &   110    \cr
11 & 1.060 &       92 & \nodata &  6900    \cr
12 & 2.800 &      160 &    6700 & 24000    \cr
13 & 1.512 &       92 & \nodata &  5800    \cr
14 & 3.268 &     1400 & \nodata & 18000    \cr
\sidehead{CDF-N}
 0 & 2.000 &   77     &   340 & 12000    \cr
 1 & 2.000 &  390     &   320 & $<4900$  \cr
 2 & 0.845 &   13     &   160 &  1800    \cr
 3 & 0.679 &   24     &   610 &   810    \cr
 4 & 1.022 &   45     &  1800 & $<1000$  \cr
 5 & 2.000 &   46     &   430 & $<4900$  \cr
 6 & 1.014 &  7.1     &    45 &  2300    \cr
 7 & 2.000 &   70     &   280 & $<4900$  \cr
 8 & 2.590 &  530     & 38000 & $<8900$  \cr
 9 & 0.761 &   12     &    38 &  $<520$  \cr
10 & 2.000 &   35     &   610 &  $<4900$ \cr
11 & 1.013 &  9.6     &    46 &  3600    \cr
12 & 0.848 &  7.5     &   260 &  $<670$  \cr
13 & 2.005 &   60     &  1600 & 19000    \cr
14 & 0.957 &   13     &  1800 &  $<880$  \cr
15 & 2.724 &  100     &  1600 & $<9900$ \cr
16 & 0.857 &  6.6     &   210 &  4400    \cr
17 & 0.961 &   94     &   400 &  7000    \cr
18 & 2.000 &   33     &   900 & $<4900$  \cr
19 & 2.000 &   56     &   720 & $<4900$  \cr
20 & 2.000 &   84     &   310 & 10000    \cr
21 & 0.678 &   13     &   210 &  $<390$  \cr
22 & 0.514 &   19     &   630 &  $<210$  \cr
23 & 0.377 & 0.66     &   110 &  $<100$  \cr
24 & 2.000 &  160     &   180 &  $<4900$ \cr
25 & 1.020 &   90     & 10000 &  $<1000$ \cr
26 & 0.475 &  1.0     &    28 &  $<170$  \cr
27 & 2.000 &   52     &   460 &  $<4900$ \cr
28 & 2.000 &   38     &  1100 &  $<4900$ \cr
29 & 1.019 &  7.2     &   200 &  $<1000$ \cr
30 & 1.147 &  120     &   240 & 20000    \cr
31 & 2.000 &   71     &   720 &  $<4900$ \cr
32 & 2.000 &   50     &   450 &  $<4900$ \cr
33 & 2.000 &   66     &  1900 & 19000 
\enddata
\label{tab6}
\end{deluxetable}

\newpage

\newpage

\newpage

\begin{figure*}[tb]
\centerline{\psfig{figure=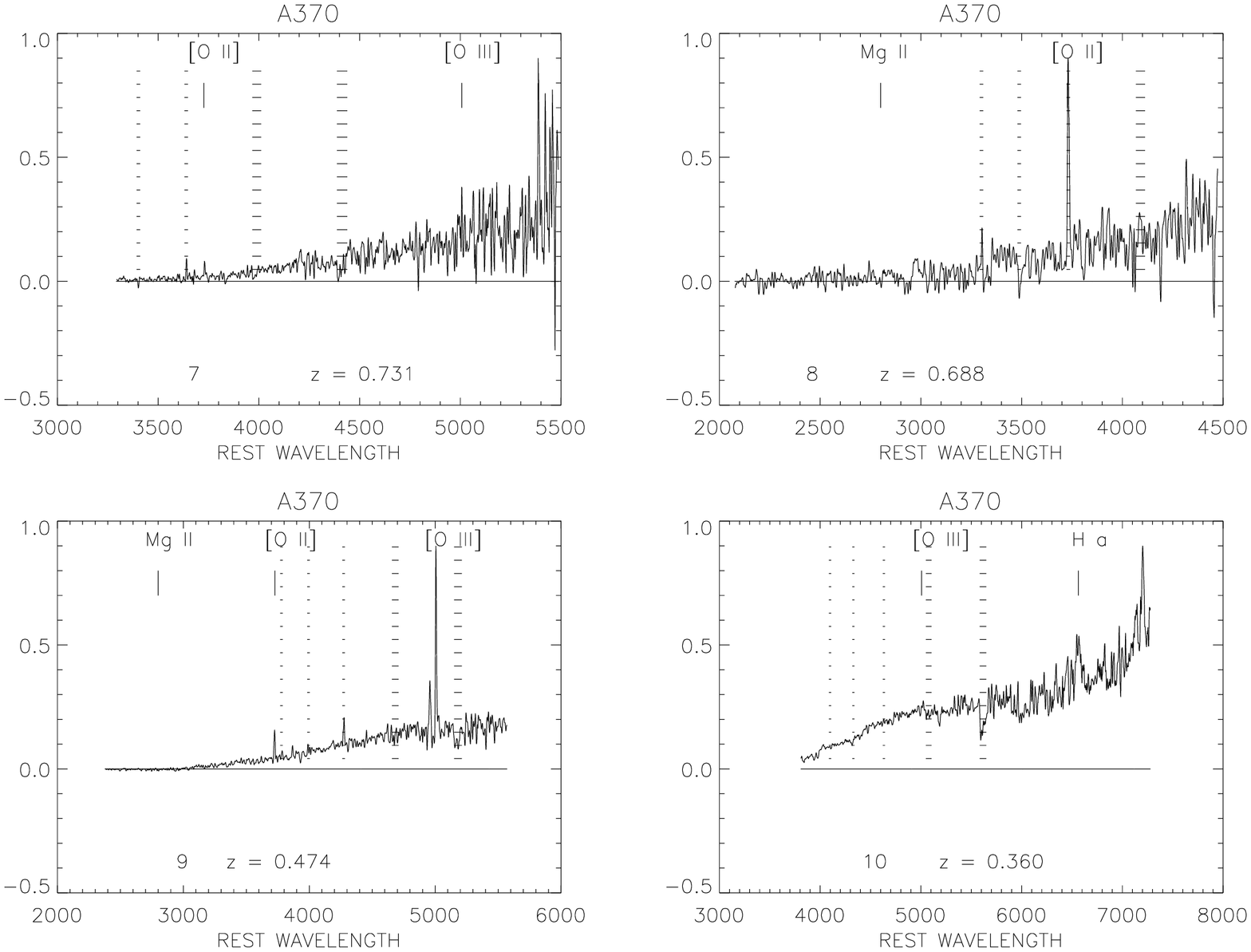,angle=90,width=7.5in}}
\figurenum{3}
\figcaption[]{
}
\end{figure*}

\newpage

\begin{figure*}[tb]
\centerline{\psfig{figure=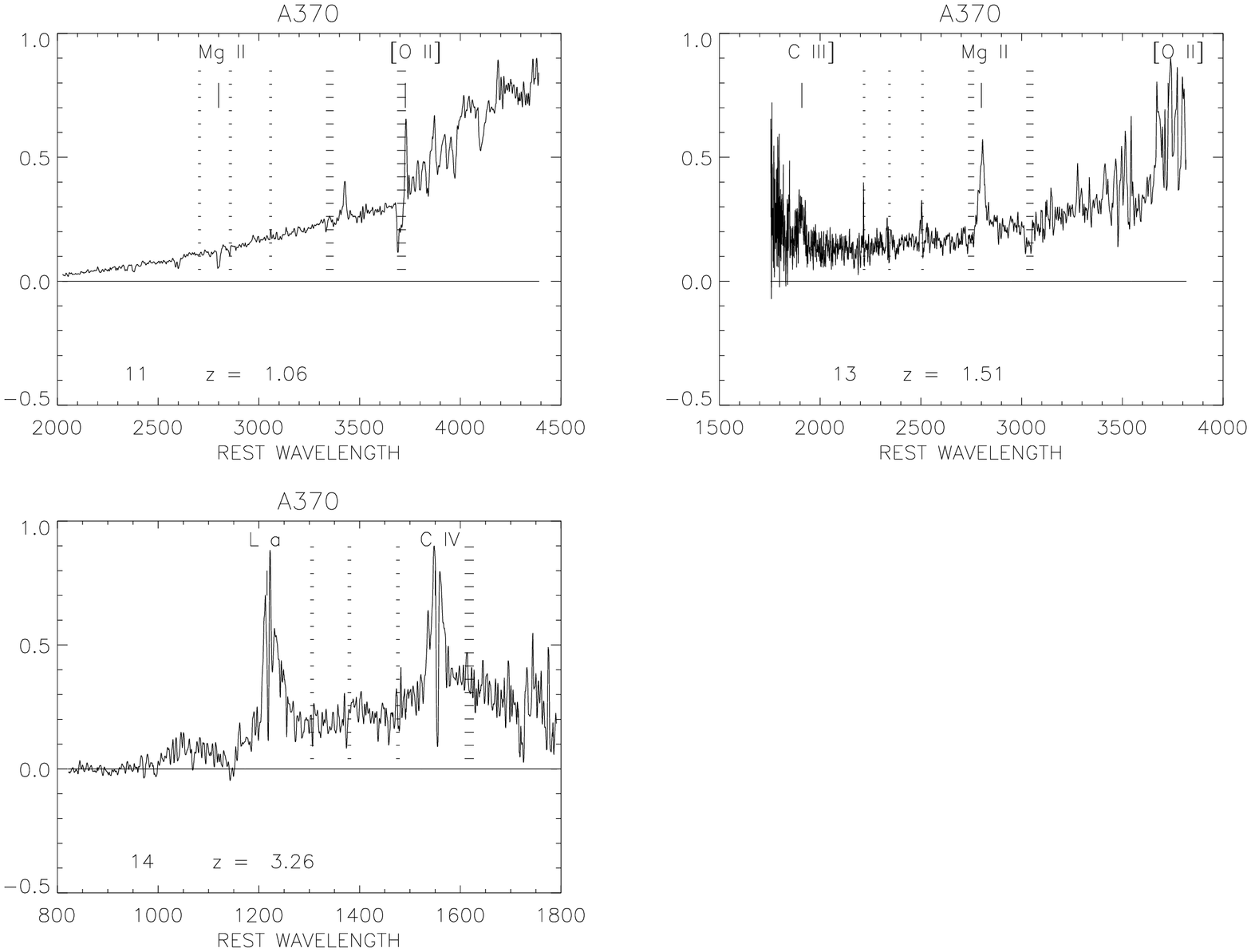,angle=90,width=7.5in}}
\figurenum{3}
\figcaption[]{
}
\end{figure*}


\begin{references}

\reference{barger99a}
Barger, A. J., Cowie, L. L., Trentham, N., Fulton, E., Hu, E. M.,
Songaila, A., \& Hall, D.\ 1999a, \aj, 117, 102

\reference{barger99b}
Barger, A. J., Cowie, L. L., Smail, I., Ivison, R. J.,
Blain, A. W., \& Kneib, J.-P.\ 1999b, \aj, 117, 2656

\reference{bcr00}
Barger, A. J., Cowie, L. L., \& Richards, E. A.\ 2000, \aj, 119, 2092

\reference{barger01a}
Barger, A. J., Cowie, L. L., Mushotzky, R. F., \& Richards, E. A.\ 2001a,
\aj, 121, 662 (B01)


\reference{bautz00}
Bautz, M. W., Malm, M. R., Baganoff, F. K., Ricker, G. R.,
Canizares, C. R., Brandt, W. N., Hornschemeier, A. E., \&
Garmire, G.P.\ 2000, \apj, 543, L119

\reference{brandt01a}
Brandt, W. N., et al.\ 2001a, \aj, in press, (astro-ph/0102411)

\reference{brandt01b}
Brandt, W. N., et al.\ 2001b, \aj, submitted

\reference{cy00}
Carilli, C. L. \& Yun, M. S.\ 2000, \apj, 530, 618

\reference{cohen00}
Cohen, J. G., Hogg, D. W., Blandford, R., Cowie, L. L., Hu, E.,
Songaila, A., Shopbell, P., \& Richberg, K.\ 2000, \apj, 538, 29 

\reference{coleman80}
Coleman, G. D., Wu, C.-C., \& Weedman, D. W.\ 1980, \apjs, 43, 393

\reference{condon92}
Condon, J. J.\ 1992, \araa, 30, 575

\reference{cowie94}
Cowie, L. L., Gardner, J. P., Hu, E. M., Songaila, A.,
Hodapp, K.-W., \& Wainscoat, R. J.\ 1994, \apj, 434, 114

\reference{cowie96}
Cowie, L. L., Songaila, A., Hu, E. M., \& Cohen, J. G.\ 1996, \aj, 112, 839

\reference{csb99}
Cowie, L. L., Songaila, A., \& Barger, A. J.\ 1999, \aj, 118, 603

\reference{cowie01}
Cowie, L. L., et al.\ 2001, \apj, 551, L9

\reference{crawford01}
Crawford, C. S., Fabian, A. C., Gandhi, P., Wilman, R. J., \&
Johnstone, R. M.\ 2001, \mnras, submitted, (astro-ph/0005242)

\reference{ferr00}
Ferrarese, L. \& Merritt, D.\ 2000, \apj, 539, L9

\reference{garmire01}
Garmire, G. P., et al.\ 2001, \apj, submitted

\reference{gebhardt00}
Gebhardt, K.\ et al.\ 2000, \apj, 539, L13

\reference{geller97}
Geller, M. J., et al.\ 1997, \aj, 114, 2205

\reference{giacconi00}
Giacconi, R.,\ et al.\ 2000, \apj, in press, (astro-ph/0007240)

\reference{hasinger01}
Hasinger, G., et al.\ 2001, A\&A, 365, L45

\reference{hodapp96}
Hodapp, K.-W.\ et al.\ 1996, NewA, 1, 177

\reference{horn00}
Hornschemeier, A. E.,\ et al.\ 2000, \apj, 541, 49

\reference{horn01}
Hornschemeier, A. E.,\ et al.\ 2001, \apj, in press, 
(astro-ph/0101494) (H01)


\reference{ivison98}
Ivison, R. J., Smail, I., Le Borgne, J.-F., Blain, A. W., Kneib, J.-P.,
B{\'e}zecourt, J., Kerr, T. H., \& Davies, J. K.\ 1998, \mnras,
298, 583

\reference{kh00}
Kauffmann, G. \& Haehnelt, M.\ 2000, MNRAS, 311, 576

\reference{lilly95}
Lilly, S. J., Tresse, L., Hammer, F., Crampton, D., 
\& Le F{\`e}vre, O.\ 1995, \apj, 455, 108

\reference{lilly96}
Lilly, S. J., LeF{\`e}vre, O., Hammer, F., \& Crampton, D.\ 1996,
\apj, 460, L1

\reference{lin96}
Lin, H., Kirshner, R. P., Shectman, S. A., Landy, S. D.,
Oemler, A., Tucker, D. L., \& Schechter, P. L.\ 1996, \apj, 464, 60

\reference{mushotzky00}
Mushotzky, R. F., Cowie, L. L., Barger, A. J., \& Arnaud, K. A.\ 2000,
\nat, 404, 459

\reference{oke95}
Oke, J. B., et al.\ 1995, \pasp, 107, 375


\reference{richards00}
Richards, E. A.\ 2000, \apj, 533, 611


\reference{schmidt98}
Schmidt, M., et al.\ 1998, A\&A, 329, 495

\reference{soucail99}
Soucail, G., Kneib, J.-P., B{\'e}zecourt, J., Metcalfe, L.,
Altieri, B., \& Le Borgne, J.-F.\ 1999, A\&A, 343, L70

\reference{tozzi01}
Tozzi, P., et al.\ 2001, \apj, submitted, (astro-ph/0103014)

\reference{wandel99}
Wandel, A.\ 1999, \apj, 519, L39


\end{references}
\end{document}